\documentclass[slac_one]{revtex4}
\usepackage{graphicx}
\usepackage{fancyhdr}
\newlength{\dslashwidth}

\newcommand{\tb}{\ensuremath{\tan\beta}}

\newcommand{\bq}{\begin{equation}}
\newcommand{\eq}{\end{equation}}
\newcommand{\ba}{\begin{array}}
\newcommand{\ea}{\end{array}}
\newcommand{\bqa}{\begin{eqnarray}}
\newcommand{\eqa}{\end{eqnarray}}

\newcommand{\lnf}{{\ifmmode \Lambda^{(N_f)} \else $\Lambda^{(N_f)}$\fi}}
\newcommand{\ms}{{\ifmmode \overline{MS} \else $\overline{MS}$\fi}}
\newcommand{\dr}{{\ifmmode \overline{DR} \else $\overline{DR}$\fi}}
\newcommand{\lms}{{\ifmmode \Lambda^{(5)}_{\overline{MS}} \else $\Lambda^{(5)}_{\overline{MS}}$\fi}}
\newcommand{\lam}{{\ifmmode \Lambda \else $\Lambda$\fi}}
\newcommand{\gev}{{\ifmmode {\rm GeV} \else ${\rm GeV}$\fi}}
\newcommand{\gevc}{{\ifmmode {\rm GeV/c^2} \else ${\rm GeV/c^2}$\fi}}
\newcommand{\tev}{{\ifmmode {\rm TeV} \else ${\rm TeV}$\fi}}
\newcommand{\tevc}{{\ifmmode {\rm TeV/c^2} \else ${\rm TeV/c^2}$\fi}}
\newcommand{\lp}{{\ifmmode L^+  \else $L^+$\fi}}
\newcommand{\lm}{{\ifmmode L^-  \else $L^-$\fi}}
\newcommand{\mlp}{{\ifmmode M(L^-) \else $M(L^-)$\fi}}
\newcommand{\mlz}{{\ifmmode M(L^0) \else $M(L^0)$\fi}}
\newcommand{\lz}{{\ifmmode L^0 \else $L^0$\fi}}
\newcommand{\ev}{{\ifmmode GeV/c^2 \else $GeV/c^2$\fi}}
\newcommand{\tri}{{\ifmmode \triangleup \else $\triangleup$\fi}}
\newcommand{\unl}{{\ifmmode U_{lL^0} \else $U_{lL^0}$\fi}}\newcommand{\gL}{{\ifmmode g_L \else $g_{L}$\fi}}
\newcommand{\gR}{{\ifmmode g_R  \else $g_{R}$\fi}}
\newcommand{\gumu}{{\ifmmode \gamma^{\mu} \else $\gamma^{\mu}$\fi}}
\newcommand{\gunu}{{\ifmmode \gamma^{\nu} \else $\gamma^{\nu}$\fi}}
\newcommand{\gdmu}{{\ifmmode \gamma_{\mu} \else $\gamma_{\mu}$\fi}}
\newcommand{\gdnu}{{\ifmmode \gamma_{\nu} \else $\gamma_{\nu}$\fi}}
\newcommand{\stw}{{\ifmmode\sin^2\theta_W \else $\sin^{2}\theta_{W}$ \fi}}
\newcommand{\sws}{{\ifmmode \;\sin^2\theta_W  \else $\;\sin^{2}\theta_{W}$ \fi}}
\newcommand{\cws}{{\ifmmode \;\cos^2\theta_W  \else $\;\cos^{2}\theta_{W}$ \fi}}
\newcommand{\sw}{{\ifmmode \;\sin\theta_W  \else $\sin\theta_{W}$ \fi}}
\newcommand{\cw}{{\ifmmode \;\cos\theta_W  \else $\;\cos\theta_{W}$ \fi}}
\newcommand{\tw}{{\ifmmode \;\tan\theta_W  \else $\;\tan\theta_{W}$ \fi}}
\newcommand{\qq}{{\ifmmode q\overline{q} \else $q\overline{q}$\fi}}
\newcommand{\lR}{{\ifmmode l_R \else $l_R$\fi}}
\newcommand{\lL}{{\ifmmode l_L \else $l_L$\fi}}
\newcommand{\nt}{{\ifmmode \nu_{\tau} \else $\nu_{\tau}$\fi}}
\newcommand{\nuR}{{\ifmmode \nu_R  \else $\nu_R$\fi}}
\newcommand{\nuL}{{\ifmmode \nu_L  \else $\nu_L$\fi}}
\newcommand{\qR}{{\ifmmode g_R  \else $q_R$\fi}}
\newcommand{\qL}{{\ifmmode q_L  \else $q_L$\fi}}
\newcommand{\qRp}{{\ifmmode q_R'  \else $q_{R}$'\fi}}
\newcommand{\qLp}{{\ifmmode q_L'  \else $q_{L}$'\fi}}
\newcommand{\est}{{\ifmmode e^{\bf \ast} \else $e^{\bf \ast}$\fi}}
\newcommand{\lst}{{\ifmmode l^{\bf \ast} \else $l^{\bf \ast}$\fi}}
\newcommand{\must}{{\ifmmode \mu^{\bf \ast} \else $\mu^{\bf \ast}$\fi}}
\newcommand{\taust}{{\ifmmode \tau^{\bf \ast} \else $\tau^{\bf \ast}$ \fi}}
\newcommand{\pperp}{{\ifmmode p_t  \else $p_t$\fi}}
\newcommand{\et}{{\ifmmode E_t  \else $E_t$\fi}}
\newcommand{\xt}{{\ifmmode x_t  \else $x_t$\fi}}
\newcommand{\smumu}{{\ifmmode \sigma_{\mu\mu}  \else $\sigma_{\mu\mu}$ \fi}}
\newcommand{\eg}{{\ifmmode e\gamma  \else $e\gamma$\fi}}
\newcommand{\epem}{{\ifmmode e^+e^-  \else $e^+e^-$\fi}}
\newcommand{\lplm}{{\ifmmode L^+L^-  \else $L^+L^-$\fi}}
\newcommand{\pp}{{\ifmmode p\overline p  \else $p\overline p$\fi}}
\newcommand{\llz}{{\ifmmode L^0\overline{L}^0 \else $L^0\overline{L}^0$\fi}}
\newcommand{\epemt}{{\ifmmode e^+e^- \to  \else $e^+e^- \to$\fi}}
\newcommand{\eb}{{\ifmmode E_{beam}  \else $E_{beam}$\fi}}
\newcommand{\ip}{{\ifmmode pb^{-1}  \else $pb^{-1}$\fi}}
\newcommand{\upm}{{\ifmmode ^{\pm}  \else $^{\pm}$\fi}}
\newcommand{\de}{{\ifmmode ^{\circ}  \else $^{\circ}$ \fi}}
\newcommand{\appr}{{\ifmmode \sim \else $\sim$ \fi}}
\newcommand{\corresp}{{\ifmmode \stackrel{\wedge}{=} \else $\stackrel{\wedge}{=}$ \fi}}
\newcommand{\sqrts}{{\ifmmode \sqrt{s} \else $\sqrt{s}$\fi}}
\newcommand{\zz}{{\ifmmode Z^0  \else $Z^0$\fi}}
\newcommand{\mz}{{\ifmmode M_{Z}  \else $M_{Z}$\fi}}
\newcommand{\mzs}{{\ifmmode M_{Z}^2  \else $M_{Z}^2$\fi}}
\newcommand{\mw}{{\ifmmode M_{W}  \else $M_{W}$\fi}}
\newcommand{\mws}{{\ifmmode M_{W}^2  \else $M_{W}^2$\fi}}
\newcommand{\mh}{{\ifmmode M_{Higgs}  \else $M_{Higgs}$\fi}}
\newcommand{\gt}{{\ifmmode \Gamma_{tot} \else $\Gamma_{tot}$\fi}}
\newcommand{\msusy}{{\ifmmode M_{SUSY}  \else $M_{SUSY}$\fi}}
\newcommand{\msusys}{{\ifmmode M_{SUSY}^2  \else $M_{SUSY}^2$\fi}}
\newcommand{\su}{{\ifmmode SU(3)_C\otimes\- SU(2)_L\otimes\- U(1)_Y \else $SU(3)_C\otimes SU(2)_L\otimes U(1)_Y$\fi}}
\newcommand{\suthree}{{\ifmmode SU(3)_C  \else $SU(3)_C$\fi}}
\newcommand{\sutwo}{{\ifmmode  SU(2)_L\otimes U(1)_Y \else $SU(2)_L\otimes U(1)_Y$\fi}}
\newcommand{\taup} {{\ifmmode \tau_{proton} \else $\tau_{proton}$\fi}}
\newcommand{\as}{{\ifmmode \alpha_{s}  \else $\alpha_{s}$\fi}}
\newcommand{\mgut}{{\ifmmode M_{GUT}  \else $M_{GUT}$\fi}}
\newcommand{\mguts}{{\ifmmode M_{GUT}^2  \else $M_{GUT}^2$\fi}}
\newcommand{\mze} {{\ifmmode m_0        \else $m_0$\fi}}
\newcommand{\mha}{{\ifmmode m_{1/2}    \else $m_{1/2}$\fi}}
\newcommand{\mb} {{\ifmmode m_{b}    \else $m_{b}$\fi}}
\newcommand{\mt} {{\ifmmode m_{t}    \else $m_{t}$\fi}}
\newcommand{\mts} {{\ifmmode m_{t}^2    \else $m_{t}^2$\fi}}

\newcommand{\mtau}{{\ifmmode m_{\tau}  \else $m_{\tau}$\fi}}
\newcommand{\dpp}{{\ifmmode \delta_{pert} \else $\delta_{pert}$\fi}}
\newcommand{\dnp}{{\ifmmode\delta_{non-pert}\else$\delta_{non-pert}$\fi}}
\newcommand{\dew}{{\ifmmode \delta_{\rm EW}\else $\delta_{\rm EW}$\fi}}
\newcommand{\rt}{{\ifmmode R_{\tau}  \else $R_{\tau} $\fi}}
\newcommand{\rz}{{\ifmmode R_{Z}  \else $R_{Z} $\fi}}

\newcommand{\swb}{{\ifmmode \sin^2\theta_{\overline{MS}} \else $\sin^2\theta_{\overline{MS}}$\fi}}
\newcommand{\cwb}{{\ifmmode \cos^2\theta_{\overline{MS}} \else $\cos^2\theta_{\overline{MS}}$\fi}}

\newcommand{\mzero}{\rm m_0}
\newcommand{\mhalf}{\rm m_{1/2}}

\begin{document}

\title{{\small{2005 International Linear Collider Workshop - Stanford,
U.S.A.}}\\ %% Please keep this conference title here
\vspace{12pt}
Dark Matter  visible by the EGRET Excess of Diffuse Galactic Gamma 
Rays?} %% Paper title goes here

\author{W. de Boer}
\affiliation{IEKP, University of Karlsruhe, Germany}

\begin{abstract}
 The public data from the EGRET space telescope on diffuse
galactic gamma rays in the energy range from 0.1 to 10 GeV
show an
excess for energies above 1 GeV in comparison with the expectations
from conventional galactic models. This excess shows
 all the key features of  Dark Matter
Annihilation (DMA), like being observable in al sky directions with
a shape corresponding to a WIMP mass between 50 and 100 GeV. The
intensity of the excess in various directions can be used to
reconstruct the DM profile, which - combined with the distribution
of visible matter - allows to calculate the rotation curve of our
Galaxy. Its peculiar shape, which is not flat, but shows a minimum
and maximum, is indeed reconstructed from the gamma rays, thus
proving that the EGRET excess traces the DM. Furthermore, the
spectral shape of the excess is consistent with mSUGRA and the WMAP
relic density for rather heavy squarks and sleptons - ${\cal{O}}(1
TeV)$ - and light charginos, neutralinos and gluinos (below 500
GeV).

\end{abstract}

\maketitle

\thispagestyle{fancy}
\section{Introduction}

If Dark Matter is a thermal relic from the early Universe, then it
is known to annihilate, since the  small amount of relic density
measured nowadays requires a large reduction in its number density.
The annihilation cross section can be obtained directly
 from its inverse proportionality to the relic density, the latter being
well known from  precision cosmology experiments \cite{wmap}. The
annihilation into quark pairs will produce $\pi^0$ mesons during the
fragmentation into hadrons, which in turn will decay into gamma
rays. Since  DM is cold, i.e. non-relativistic, the fragmenting
quarks have an initial energy equal to the WIMP mass. The gamma
spectrum from such mono-energetic quarks is well known from
electron-positron colliders, which produce exactly such states. For
heavy WIMP masses the gamma spectrum is considerably harder than the
background spectrum, mainly from $\pi^0$ mesons produced in
pp-collisions from cosmic rays with gas in the disk. Such an excess
of hard gamma rays has indeed been observed by the EGRET satellite
and the relative contributions from background and DM annihilation
signal can be obtained by fitting their different shapes with a free
normalization factor for background and signal.
 The present analysis on diffuse galactic gamma rays differs from previous
 ones - see reviews \cite{jungman,bergstrom,sumner,bertone} and references therein -
by considering simultaneously the complete sky map {\it and} the
energy spectrum, which allows us to constrain both the halo
distribution {\it and} the WIMP mass\cite{deboer,deboer1}.

\section{Indirect Dark Matter Detection}\label{sec2}

The neutral particles play  a very special role for indirect DM
searches, since they point back to the source, thus  providing a
perfect means to reconstruct the intensity (halo) profile of the DM
by observing the intensity of the gamma ray emissions in the various
sky directions. Of course, this assumes that one can distinguish
between the gamma rays from  DMA  and the ones from the background,
which is possible because of the different energy spectra: the gamma
rays from the mono-energetic quarks from DMA produce a significantly
harder spectrum than the gammas from nuclear interaction, which are
produced by the interactions between quarks with a steeply falling
power law spectrum ($\propto E^{-2.7}$).

The spectral shape of the gamma rays from either the backgrounds or
the mono-energetic quarks are well known from accelerator
experiments and can be obtained from the well-known PYTHIA code for
quark fragmentation\cite{pythia}.

A very detailed gamma ray distribution over the whole sky was
obtained by the Energetic Gamma Ray Emission Telescope EGRET, one of
the four instruments on the Compton Gamma Ray Observatory CGRO,
which collected data during nine  years, from  1991 to 2000.
  The EGRET data is  publicly available as
high resolution (0.5x0.5 degree) sky maps from the NASA
archive\footnote{NASA archive: http://cossc.gsfc.nasa.gov/archive/.}
with detailed information how to analyse them.

\begin{figure}
\begin{center}
 \includegraphics [width=0.32\textwidth,clip]{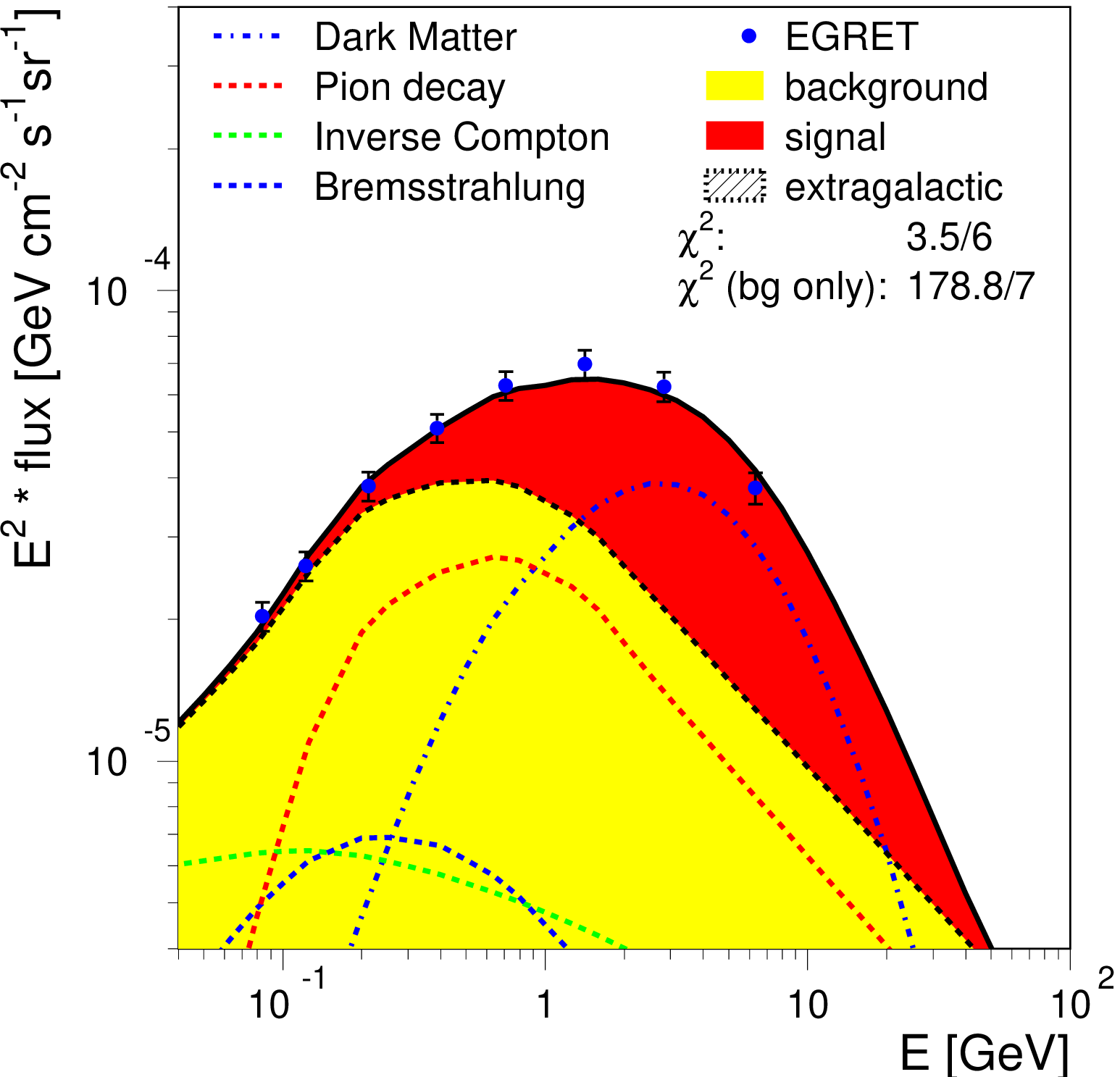}
 \includegraphics [width=0.32\textwidth,clip]{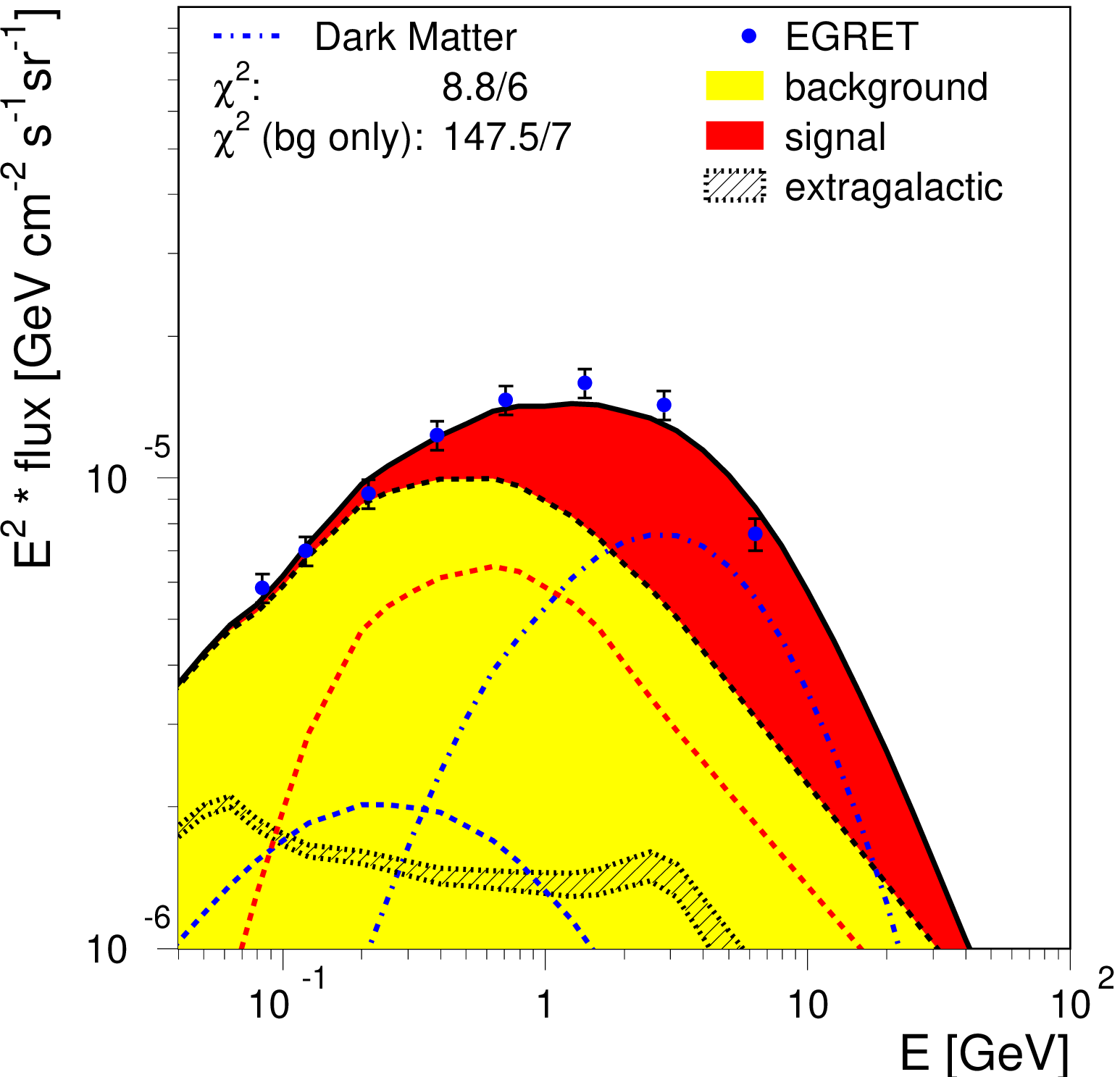}
 \includegraphics [width=0.32\textwidth,clip]{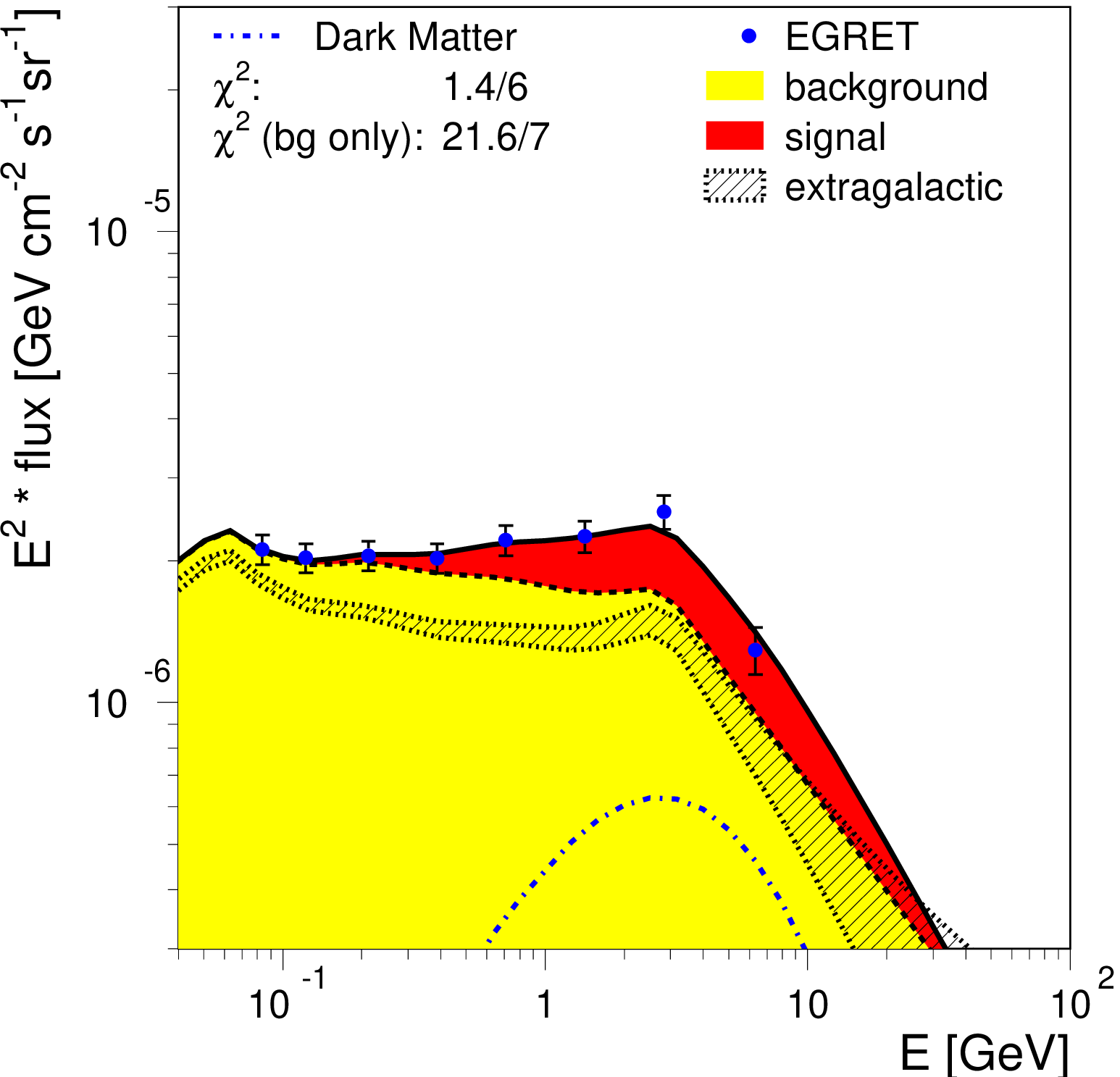}
 \caption[]{
 The diffuse gamma-ray energy spectrum of 3 angular regions: from left to right:
  towards the galactic centre (latitudes  $0^\circ<|b|<5^\circ$; longitudes $0^\circ<|l|<30^\circ$),
   the galactic anticentre ($0^\circ<|b|<10^\circ$;  $90^\circ<|l|<270^\circ$)and the pole regions
   ($60^\circ<|b|<90^\circ$;  $0^\circ<|l|<360^\circ$),
  as measured by the EGRET space telescope.
The  total background (DMA)  is
   indicated by the light (yellow)   (dark (red))  shaded area, respectively.
   The various contributions to the
background and the DMA contribution are indicated as well.
 \label{excess}}
\end{center}
\end{figure}

\begin{figure}
\begin{center}
 \includegraphics [width=0.45\textwidth,clip]{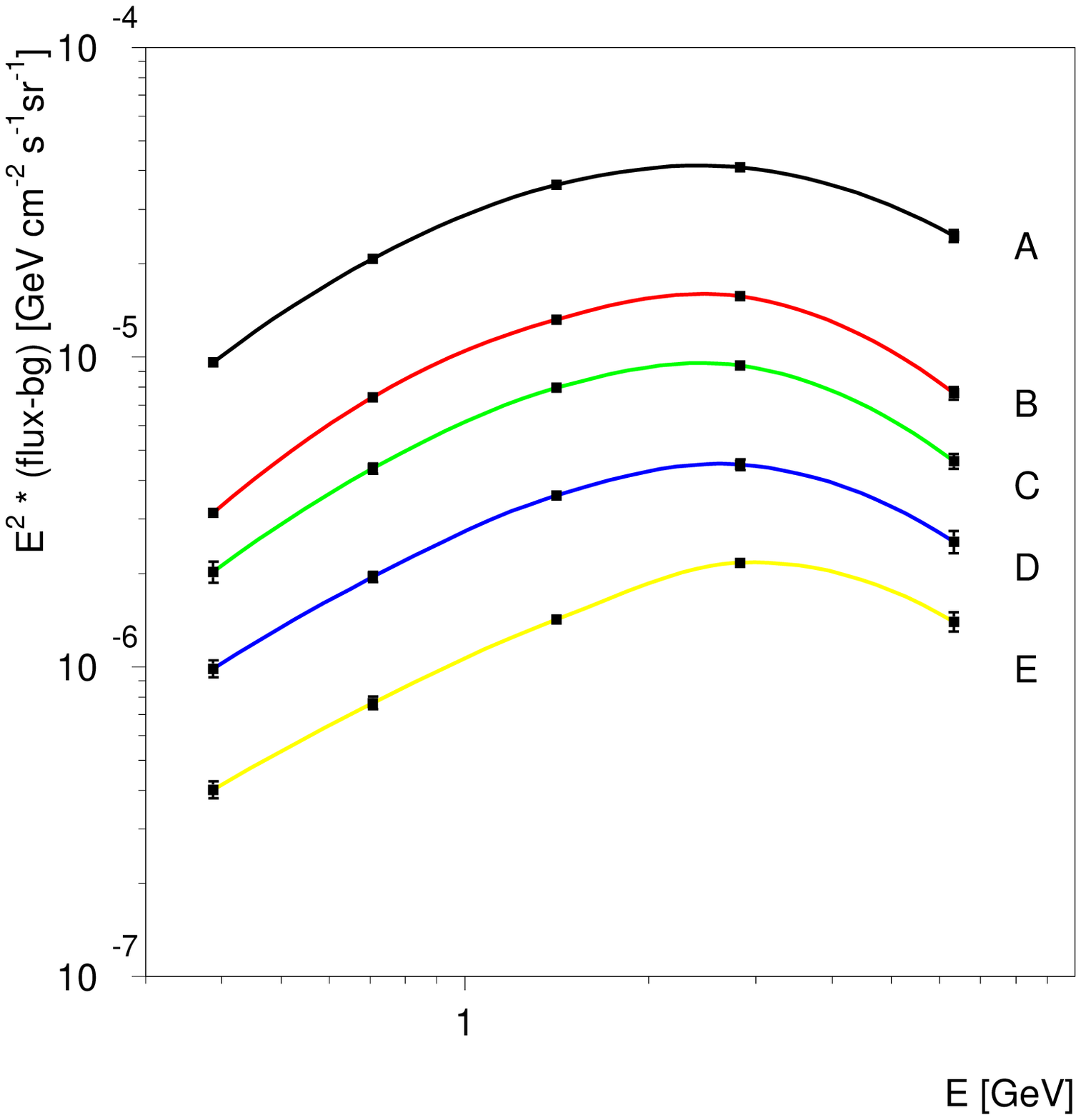}
\includegraphics [width=0.45\textwidth,clip]{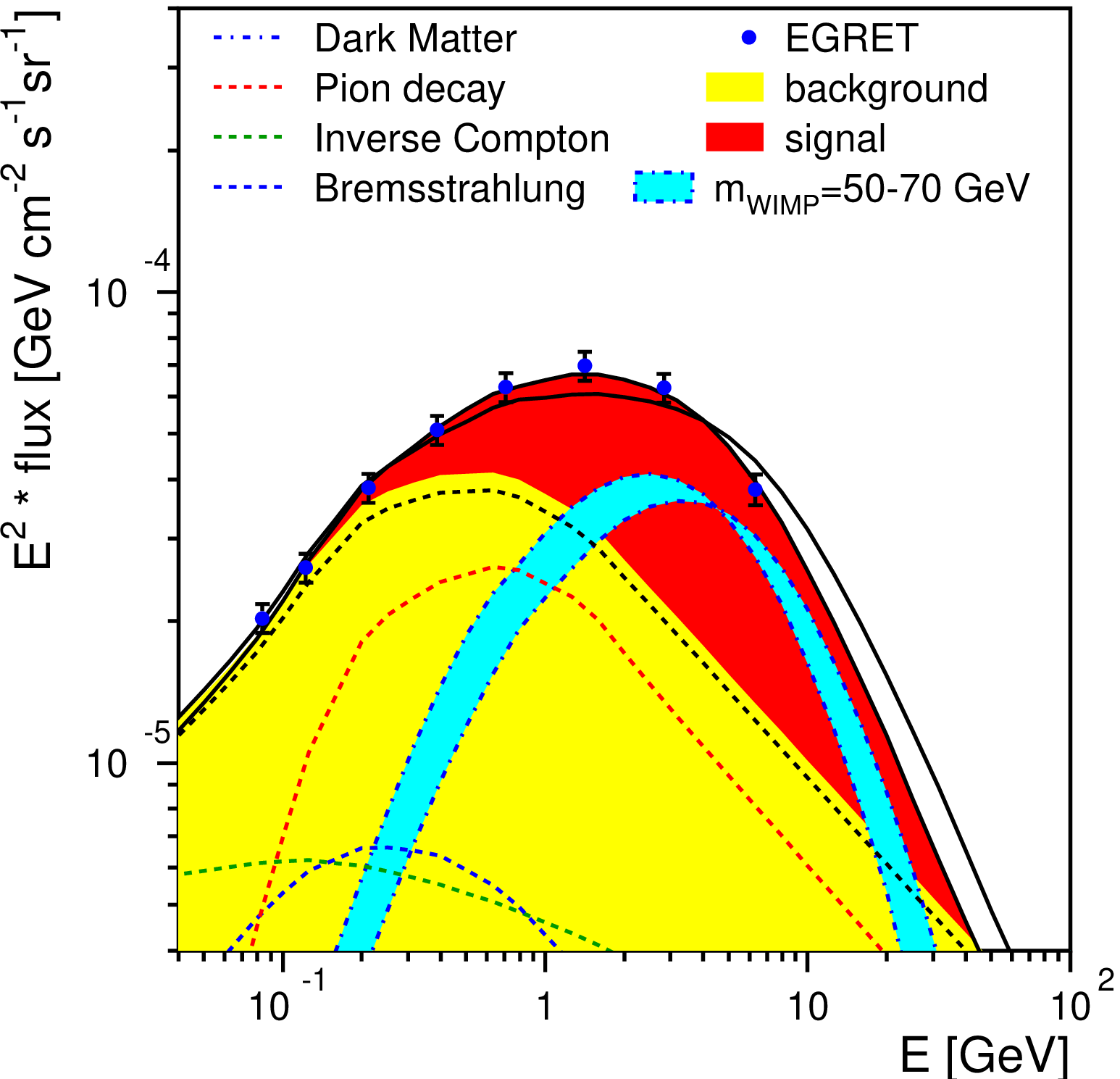}
 \caption[]{
Left: the energy spectrum of the EGRET excess, defined as the
difference between the data and the fitted background contribution
for 5 of the panels of Fig. \ref{excess}. Only the small statistical
errors have been plotted. Right: the medium shaded (blue) band shows
the effect of varying the WIMP mass between 50 and 70 GeV.}
 \label{diff}
\end{center}
\end{figure}

It was already noticed in 1997 that the EGRET data showed an excess
in the galactic disk\cite{hunter} of gamma ray fluxes for energies
above 1 GeV if compared with conventional galactic models and
repeated later for all sky directions\cite{optimized}.
 Fitting the  shape of the contributions from Galactic background, extragalactic background
 and DMA signal to the EGRET data yielded
 astonishingly good fits with the free normalization of the background agreeing reasonably
 well with the absolute predictions  of the conventional GALPROP model\cite{optimized}
 for the energies between 0.1 and 0.5 GeV. Above these energies a clear contribution from Dark
Matter annihilation is needed,  but the excess in different sky
directions can be explained by a single WIMP mass around 60 GeV  and
a single boost factor, as shown in Fig. \ref{excess} for three
different sky directions. The high quality of the EGRET data can be
appreciated from Fig. \ref{diff}, where also the sensitivity to the
WIMP mass has been plotted.

\begin{figure}
\begin{center}
 \includegraphics [width=0.32\textwidth,clip]{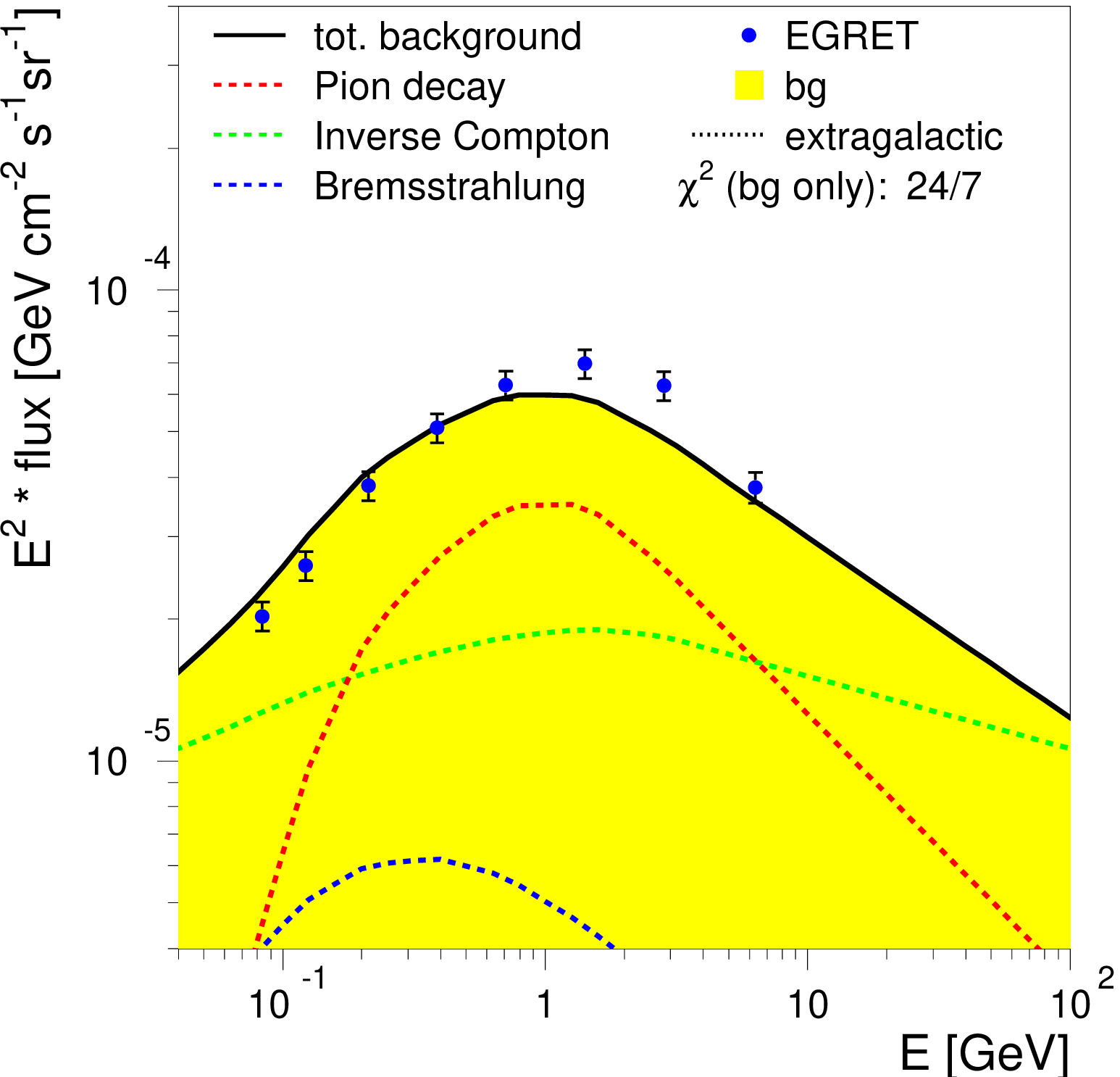}
 \includegraphics [width=0.32\textwidth,clip]{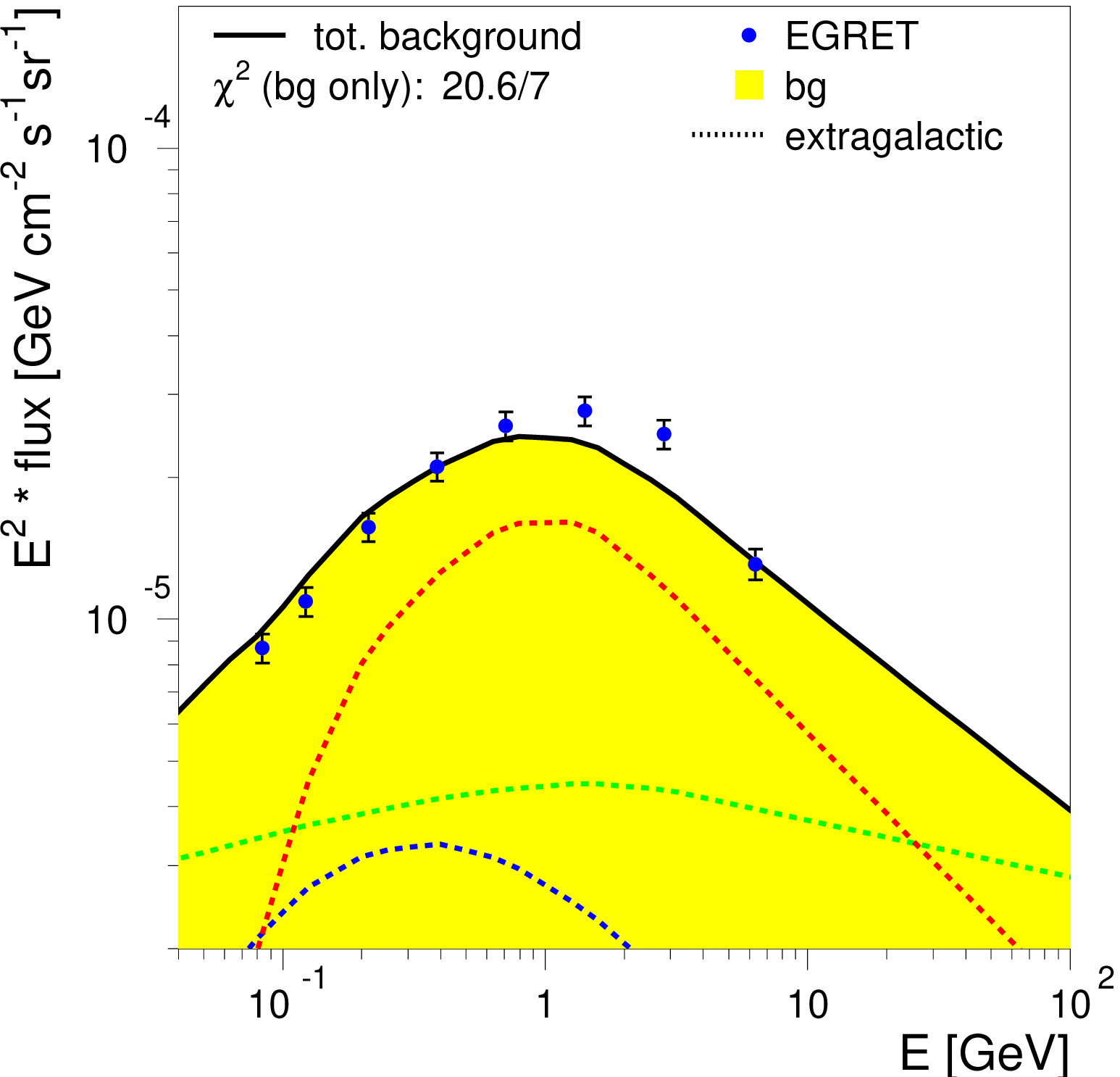}
 \includegraphics [width=0.32\textwidth,clip]{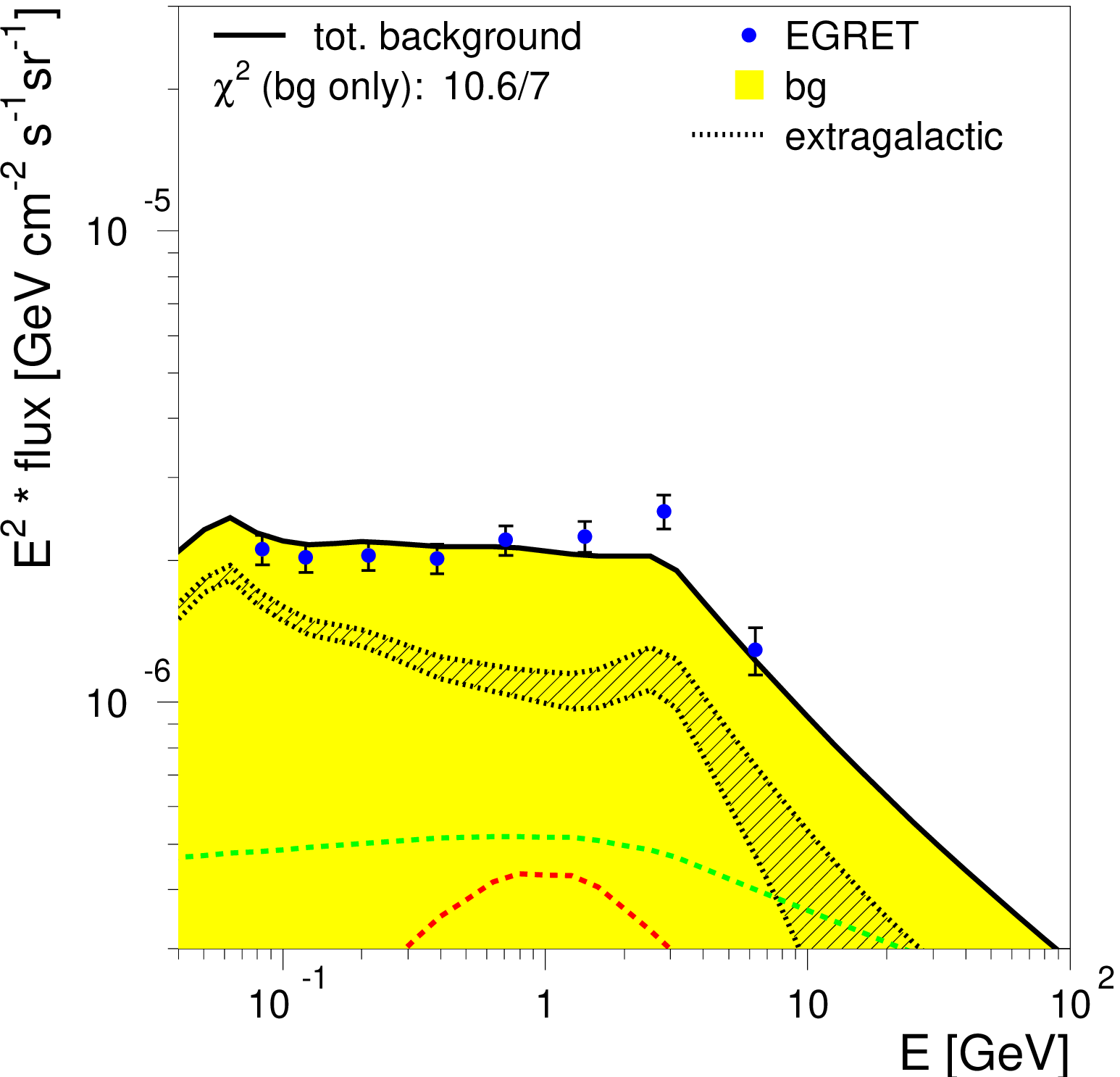}
 \caption[]{
As in Fig. \ref{excess}, but without DMA and the background shape
from Ref. \cite{optimized}, which is ``optimized'' to explain the
EGRET excess without DM. The $\chi^2$ is still poor, as indicated in
the figure.
 \label{excess1}}
\end{center}
\end{figure}

Alternative explanations for the excess have been plentiful.
 A summary of these discussions has been given in Ref.
\cite{optimized}, where it was noted that by increasing the
intensity of the electron and proton spectra at high energies in
comparison with the locally measured ones, the description of the
data can be improved.
The problem with this "optimized solution" is however that the shape
of the gamma spectra is  still not   reproduced well, as shown in
Fig. \ref{excess1}. But it is exactly the shape, which was well
measured by EGRET, because the relative errors between neighbouring
energies are roughly half of the normalization error of 15\%.  The
probability of this optimized solution is below 10$^{-7}$. Adding DM
to the optimized model improves the fit probability to 0.8. Of
course, the DM contribution is smaller than in case of the
conventional background in Fig. \ref{excess}, which results in a
reduction of the DM normalization factor by roughly a factor three.
As with the optimized model, the absolute prediction of the shape
proposed by Kamae et al. \cite{kamae} overshoots the low energy data
and undershoots the high energy data, so if only the shape is fitted
with a free normalization factor, the excess is always present.

An alternative way of formulating the problems of  models without
DMA: if the shape of the EGRET excess can be explained perfectly in
all sky directions by a gamma contribution originating from the
fragmentation of mono-energetic quarks, it is very difficult to
replace such a contribution by an excess from nuclei (quarks) (or
electrons) with a steeply falling energy spectrum. In addition, the
spatial distribution of gamma rays from DMA is different from the
gamma rays from the background.

Although the DM annihilation cross section is known from the relic
density, this does not mean we can calculate the flux from DMA,
since the flux can be enhanced (``boosted'') by the clustering of DM
and the cross section can be lowered by the reduced   temperature in
the present universe (when the p-wave contributions dominate).
Therefore we have to keep the absolute normalization of the DMA flux
as a free parameter. The different fluxes for the different sky
directions in Fig. \ref{excess} in principle yield information on
the halo profile. To obtain it more precisely from a fit  requires
first of all a much finer sampling of the sky, which is no problem
with the high statistics from EGRET (see Fig. \ref{diff}) and
secondly that one has a functional form for the profile. A survey of
the optical rotation curves of 400 galaxies shows that the halo
distributions of most of them can be fitted either with the cuspy
Navarro-Frenk-White (NFW) or the cored pseudo-isothermal profile or
both \cite{jimenez}. These halo profiles can be parametrized as
follows:
\begin{equation}
 \rho (r)=\rho_0\cdot (\frac{r}{a})^{-\gamma}\left[1+
 (\frac{r}{a})^\alpha\right] ^{\frac{\gamma-\beta}{\alpha}},
\label{profile0}
 \end{equation}
 where $a$ is a scale radius and the slopes $\alpha$, $\beta$ and
 $\gamma$ can be roughly thought of as the radial dependence at $r\approx a$,
 $r>>a$ and $r<<a$, respectively.
The cuspy NFW profile \cite{nfw} is defined by  $(\alpha, \beta,
\gamma)$ =(1,3,1) for a scale $a=10$ kpc. The isothermal profile
with $(\alpha, \beta, \gamma)$ =(2,2,0) has no cusp ($\gamma=0$),
but a core which is taken to be the size of the inner Galaxy, i.e.
$a=5$ kpc and $\beta=2$ implies a flat rotation curve. The EGRET
data towards the Galactic center do not show a strong cusp, so we
use  a cored isothermal halo.

\begin{figure}
\begin{center}
\includegraphics [width=0.35\textwidth,clip]{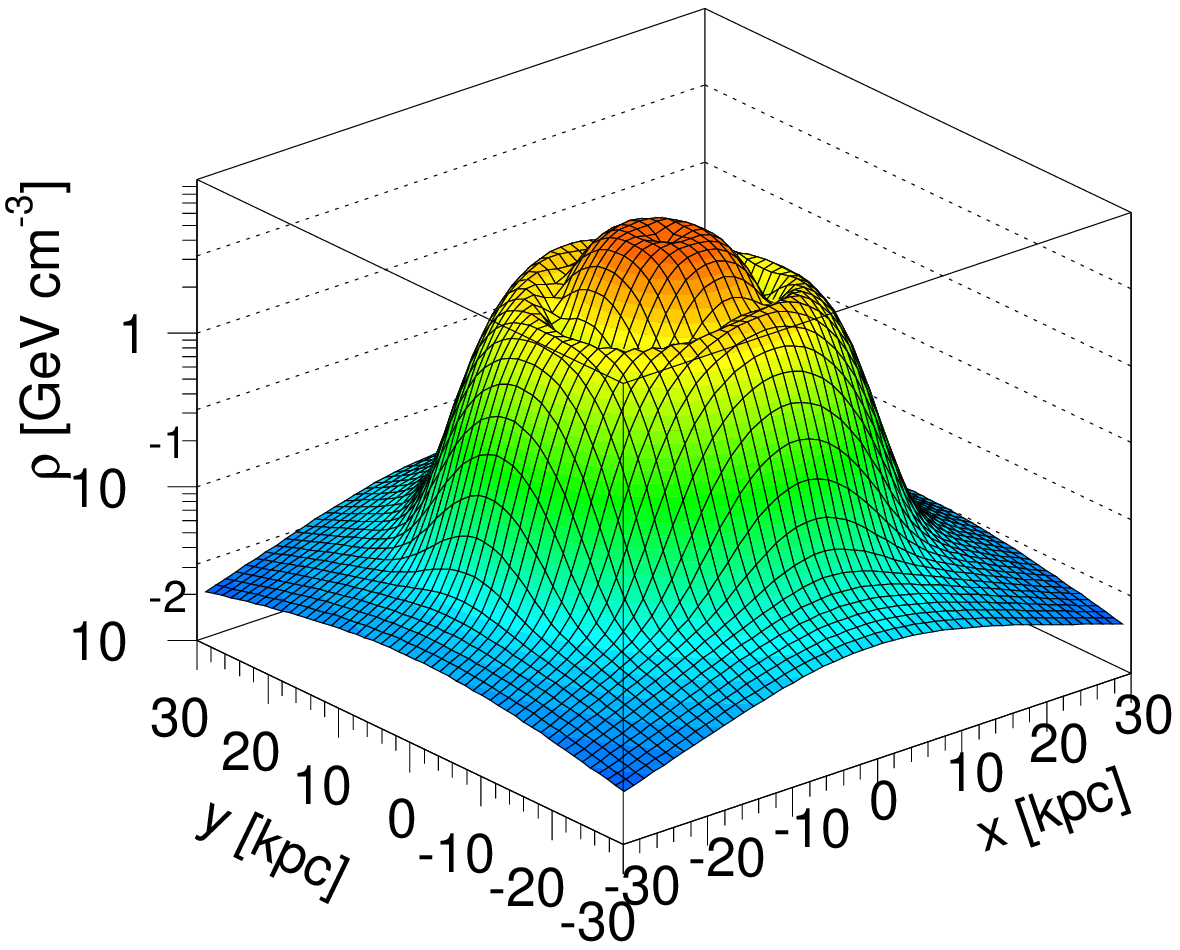}
\includegraphics [width=0.35\textwidth,clip]{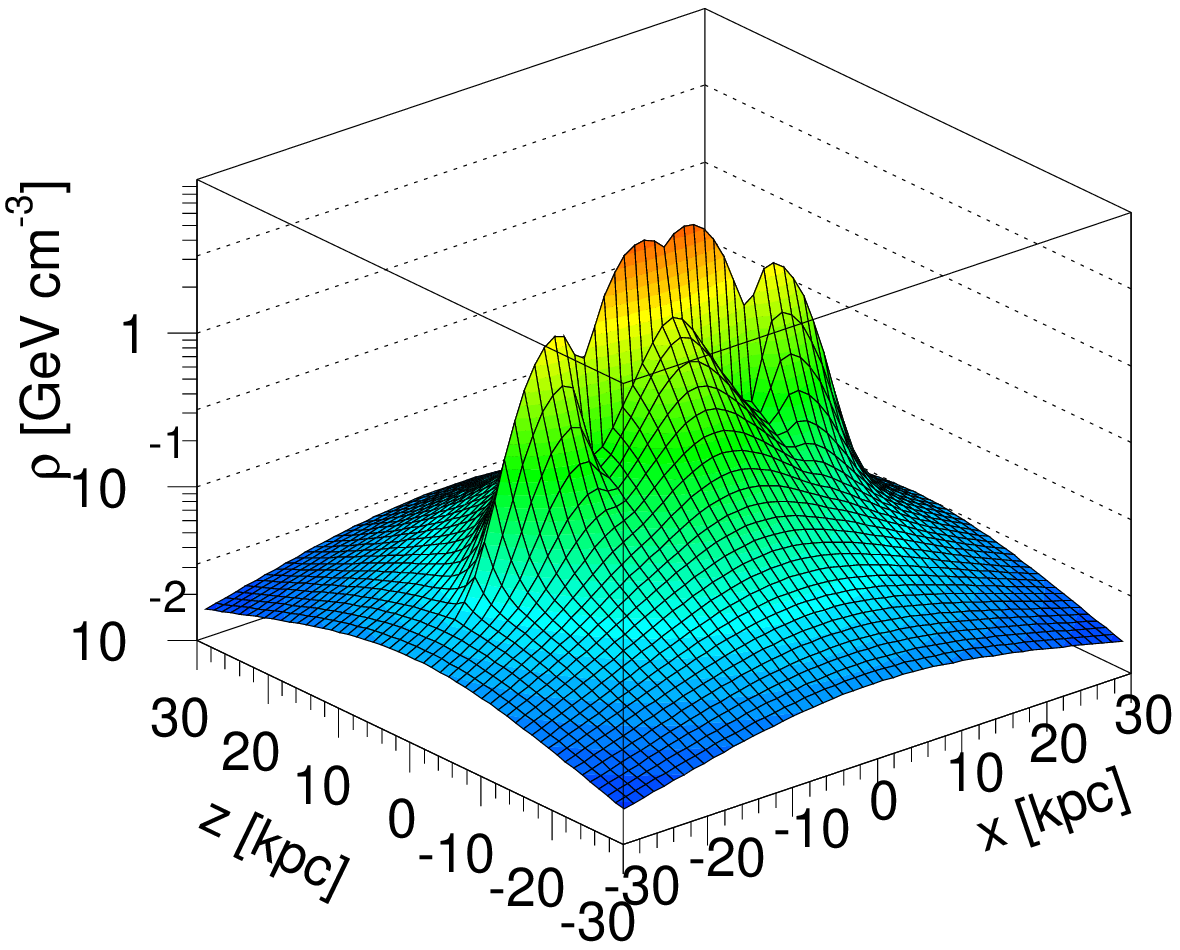}
 \caption[]{  3D-distributions of the  DM halo profile
  in the galactic xy-plane   and xz-plane.
  In the disk (xy-plane) the vertical
axis shows clearly the enhancement in density at a radius of 14 kpc,
while  the xz-plane,  going along the x-axis in the disk (z=0) shows
the effect of both rings. }
 \label{profile}
\end{center}
\end{figure}
\begin{figure}
\begin{center}
 \includegraphics [width=0.36\textwidth,clip]{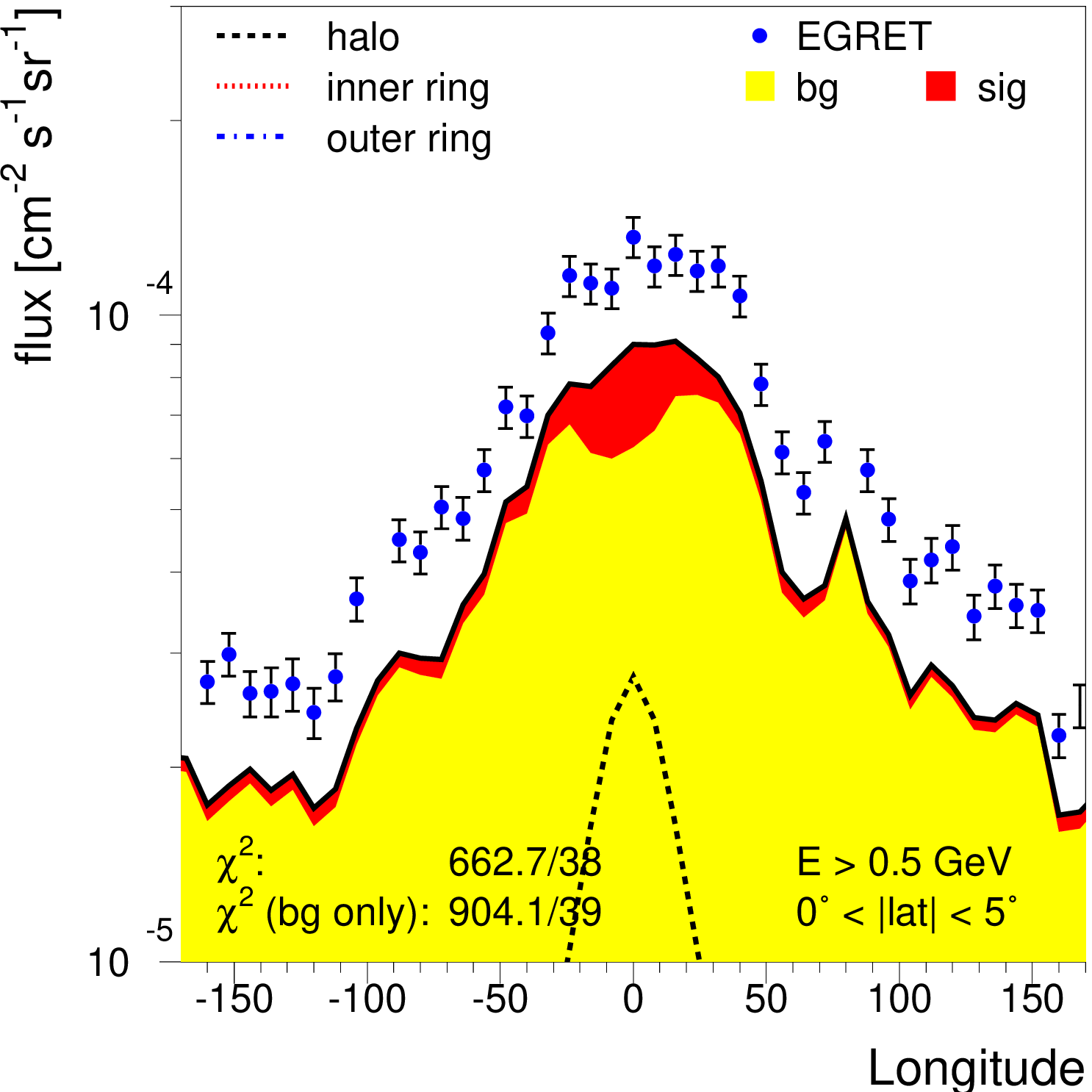}
 \includegraphics [width=0.36\textwidth,clip]{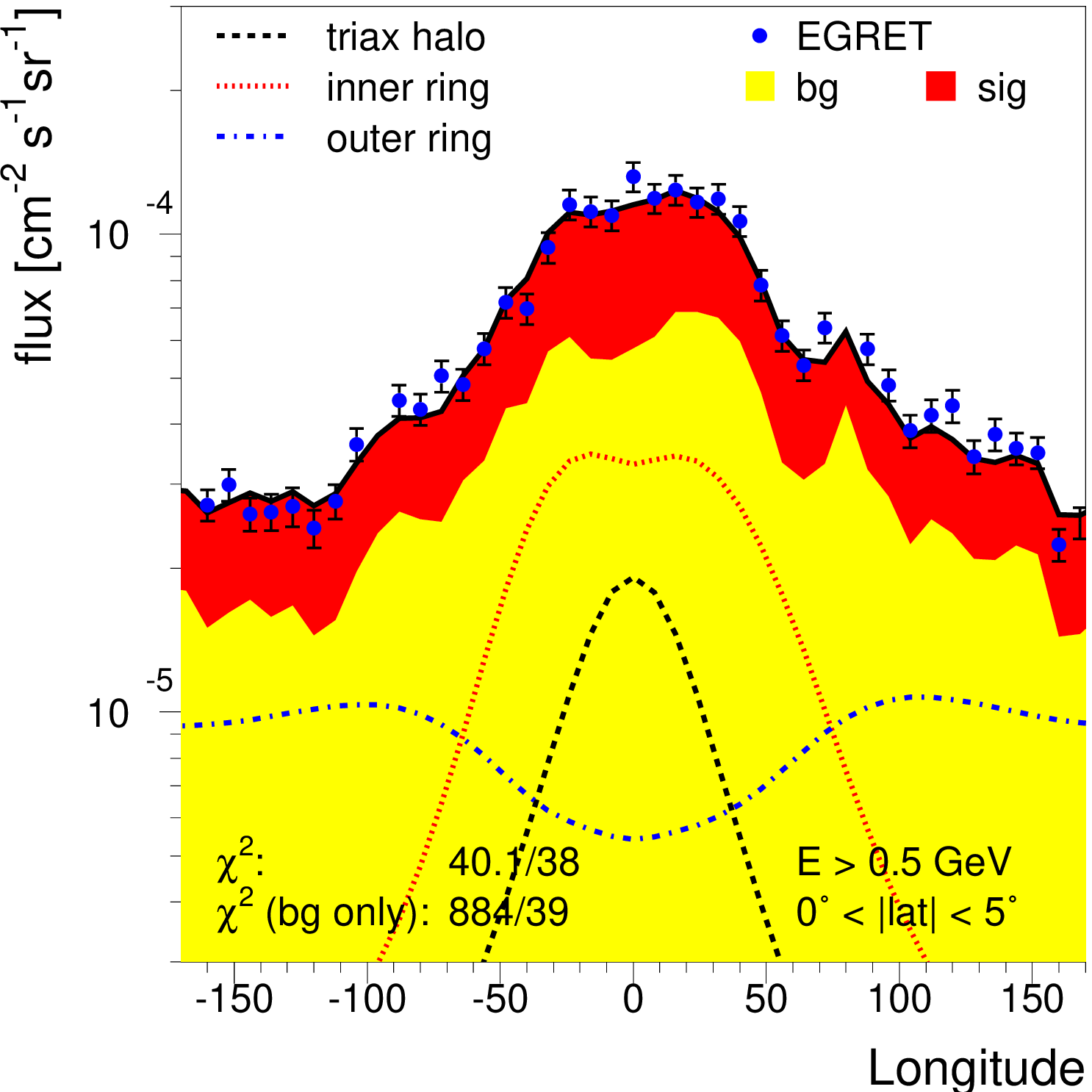}
 \includegraphics [width=0.36\textwidth,clip]{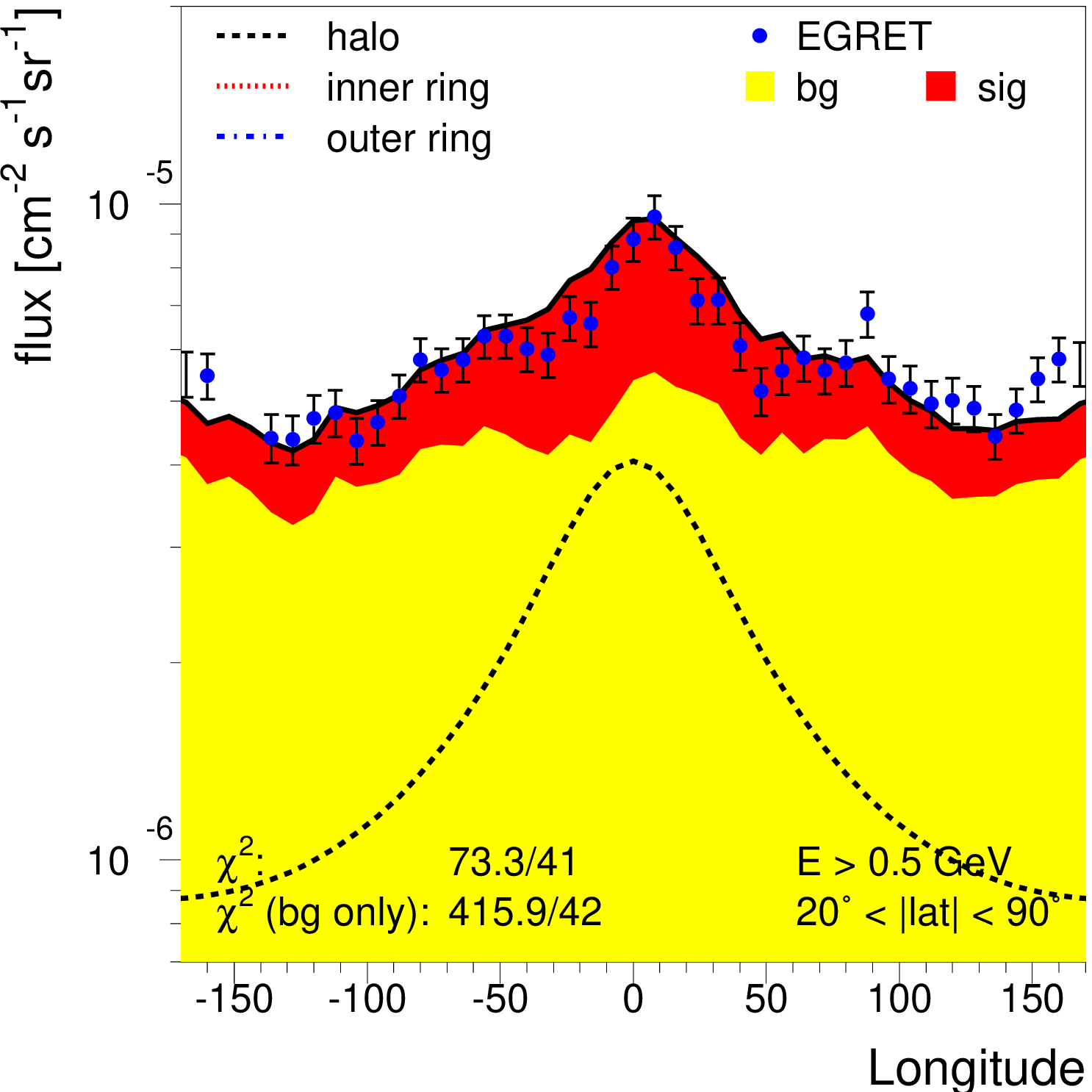}
 \includegraphics [width=0.36\textwidth,clip]{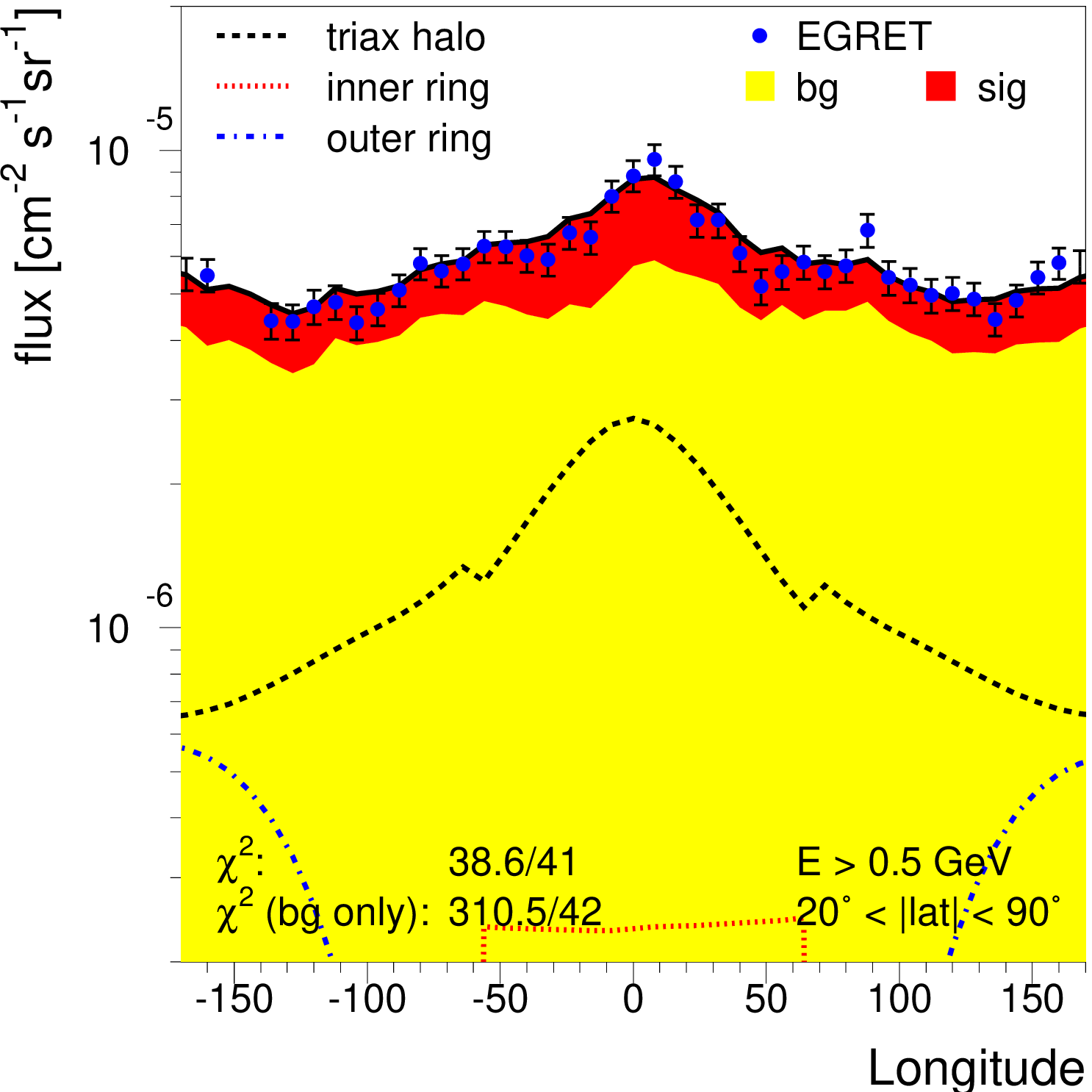}
 \caption[]{Top row:
 the longitude distribution of diffuse gamma-rays in the disk of the Galaxy
 (latitudes $0^\circ<|b|<5^\circ$)
 %and two energy bins: $E_\gamma<0.5$ GeV (left) and $E_\gamma>0.5$ GeV (right).
for the pseudo-isothermal profile without (left) and with rings
(right).
 The points represent the EGRET data.
% The contributions from the background and the neutralino annihilation signal have
%  been indicated by the light (yellow) and dark (red) shaded area, respectively.
  Bottow row: as above for the polar regions of our Galaxy (latitudes  $20^\circ<|b|<90^\circ$) .}
 \label{long}
\end{center}
\end{figure}

 Keeping the boost factor the same for all sky
directions and fitting the DM profile in 180 independent sky
directions one finds the following surprising result: in addition to
the pseudo-isothermal profile defined above the EGRET excess shows a
substructure in the form of toroidal rings at 4 and 14 kpc, as
displayed in Fig. \ref{profile}. The need for these additional rings
is most easily seen by comparing the longitudinal profiles in the
galactic plane and towards the galactic poles. As shown in Fig.
\ref{long} the pole regions are described reasonably well without
rings, but for the galactic plane the excess is dominated by the
inner ring for the inner Galaxy (longitudes $|l|< 50^\circ$ ) and by
the outer ring for the outer galaxy (longitudes $|l|>70^\circ$).
Note that for each bin only the flux integrated for data above 0.5
GeV has been plotted.

 The position and shape of the inner ring coincides
with the ring of molecular hydrogen. Molecules form from atomic
hydrogen in the presence of dust or heavy nuclei. So a ring of
neutral hydrogen suggests an attractive gravitational potential. The
position and shape of the outer ring coincides with the ring of
stars, discovered in 2002 by several independent
groups\cite{newberg,ibata,yanny,rocha-pinto1}. This ring is thought
to originate from the infall of a  galaxy, so additional DMA is
expected there.

\begin{figure}
\begin{center}
 \includegraphics [width=0.6\textwidth,clip]{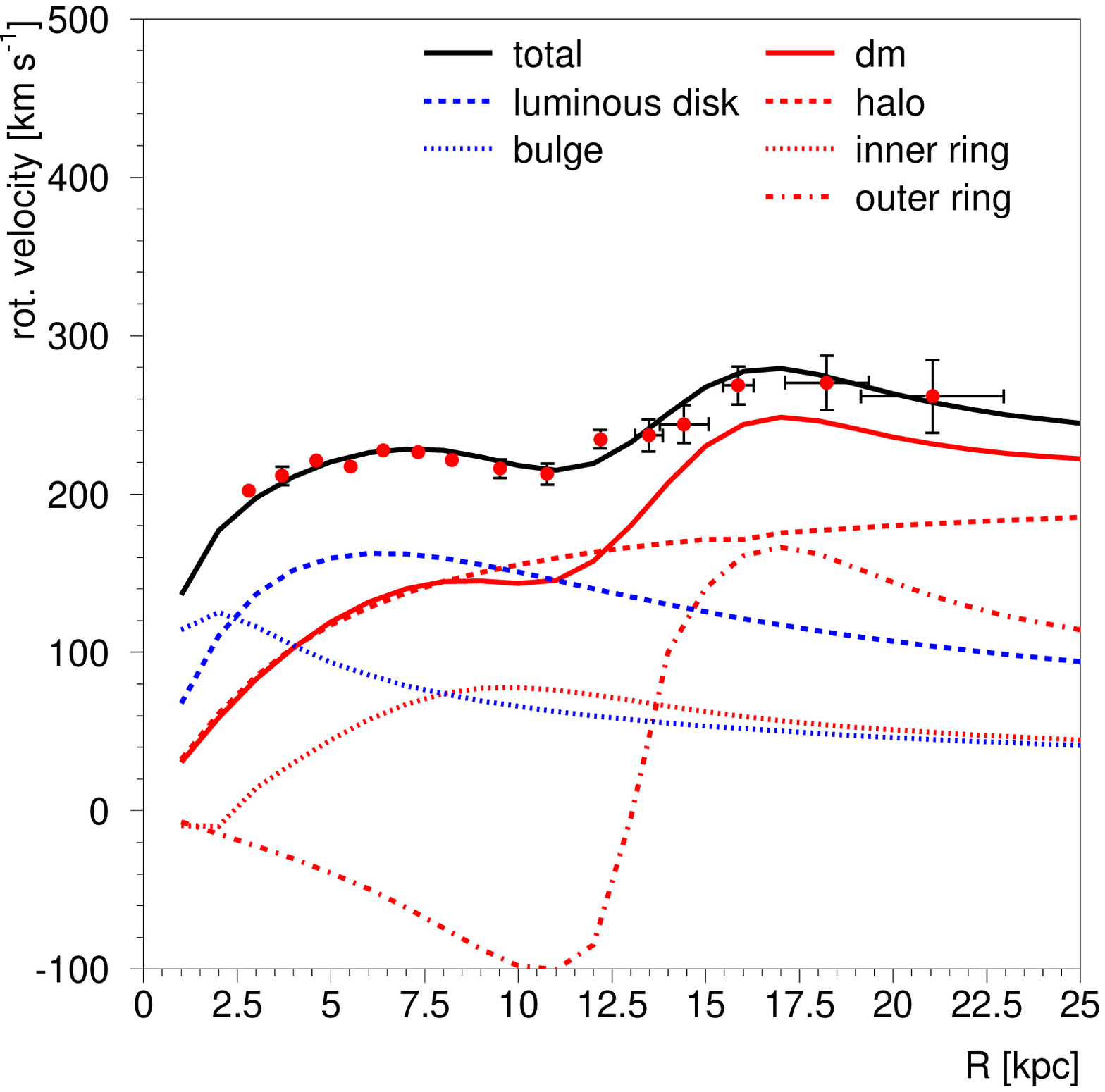}
 \caption[]{
  The rotation curve from our Galaxy with the DM contribution determined from the
   EGRET excess of diffuse gamma rays. The data are averaged from Ref. \cite{deboer}.}
 \label{rot}
\end{center}
\end{figure}

To prove that the enhanced gamma ray density is indeed connected to
non-baryonic mass the rotation curve was reconstructed from the
excess of the diffuse gamma rays in the following way: since the
flux determines the number density of DM for a given boost factor
and since the mass of each WIMP is around 60 GeV, one can determine
the mass in the ring and consequently predict the rotation
curve\footnote{For the outer ring a total DM mass of about  $\rm
9\cdot10^{10}$ solar masses was found in comparison with about $\rm
10^9$ solar masses in the form of stars.}. The two ring model
describes the peculiar change of slope at 11 kpc well, as shown in
Fig. \ref{rot}. The data was obtained as average of the data in Ref.
\cite{deboer1}. The contributions from each of the mass terms have
been
 shown separately. The basic explanation for the negative contribution from the outer ring
 is that a tracer star
 inside  the ring at 14 kpc feels an outward force from the ring, thus a negative
 contribution to the rotation velocity.
 The fact that the shape of the rotation curve can be reconstructed
 from the EGRET excess of gamma rays clearly shows that this excess
 traces the DM!

\section{Comparison with Supersymmetry}

Supersymmetry~\cite{susyrev} presupposes a symmetry between fermions
and bosons, which can be realized in nature only if one assumes each
particle with spin j has a supersymmetric partner with spin $\vert
j-1/2\vert$ ($\vert j+1/2\vert$ for the Higgs bosons). This leads to
a doubling of the particle spectrum. Obviously SUSY cannot be an
exact symmetry of nature; or else the supersymmetric partners would
have the same mass as the normal particles. The mSUGRA model, i.e.
the Minimal Supersymmetric Standard Model (MSSM) with supergravity
inspired breaking terms, is characterized by only 5 parameters:
$m_0,~m_{1/2},~\tb,~\mbox{sign}(\mu), ~A_0$. Here $m_0$ and
$m_{1/2}$ are the common masses for the gauginos and scalars at the
GUT scale, which is determined by the unification of the gauge
couplings. Gauge unification is still possible with the precisely
measured couplings at LEP~\cite{bs}. The ratio of the vacuum
expectation values of the two Higgs doublets is called \tb ~ and
$A_0$ is the trilinear coupling at the GUT scale. We only consider
the dominant trilinear couplings of the third generation of quarks
and leptons and assume also $A_0$ to be unified at the GUT scale.
The absolute value of the Higgs mixing parameter $\mu$ is determined
by electroweak symmetry breaking, while its sign is taken to be
positive, as preferred by the anomalous magnetic moment of the
muon~\cite{bs}.
\begin{figure}
\begin{center}
 \includegraphics [width=0.45\textwidth,clip]{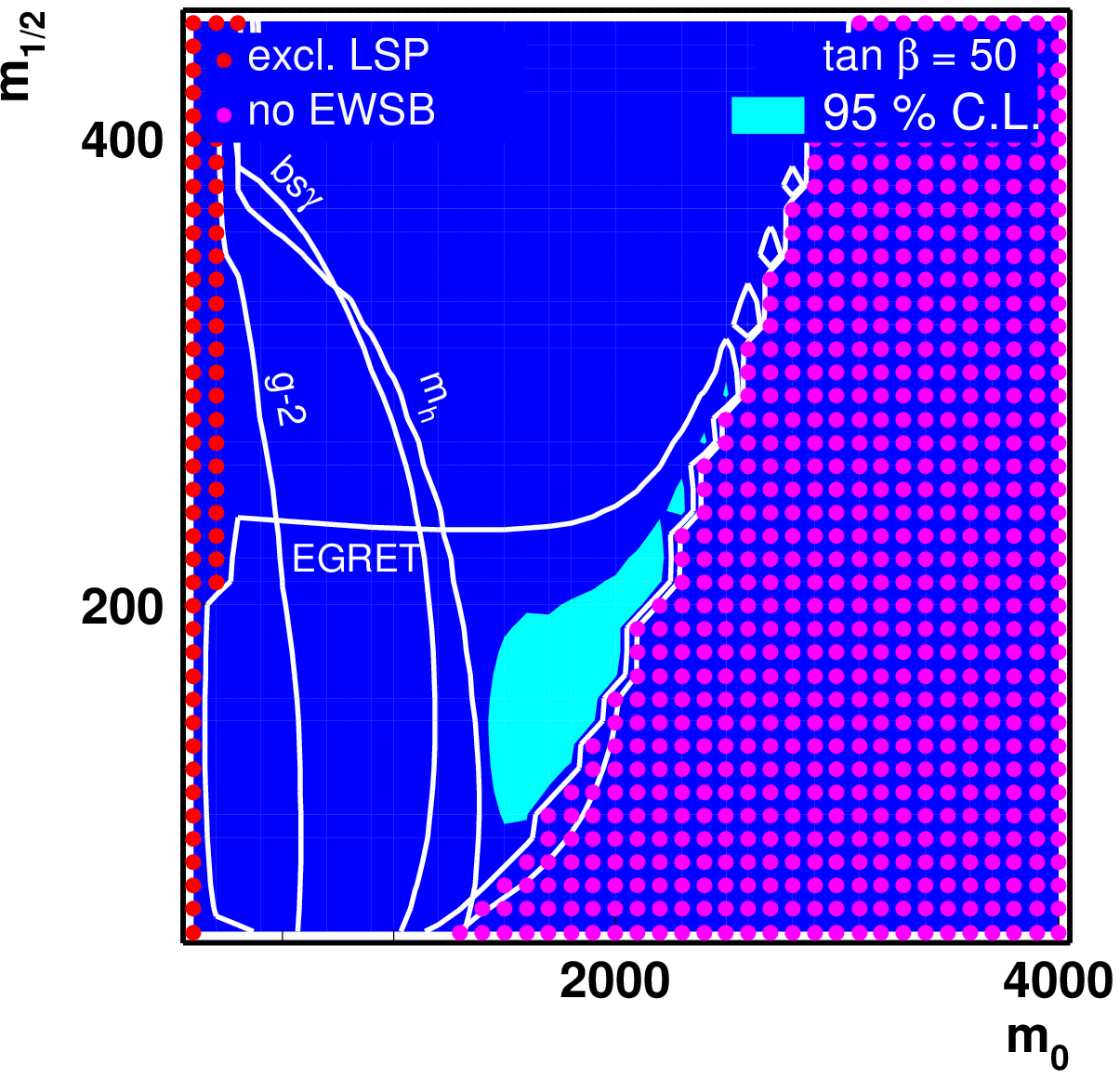}
 \includegraphics [width=0.45\textwidth,clip]{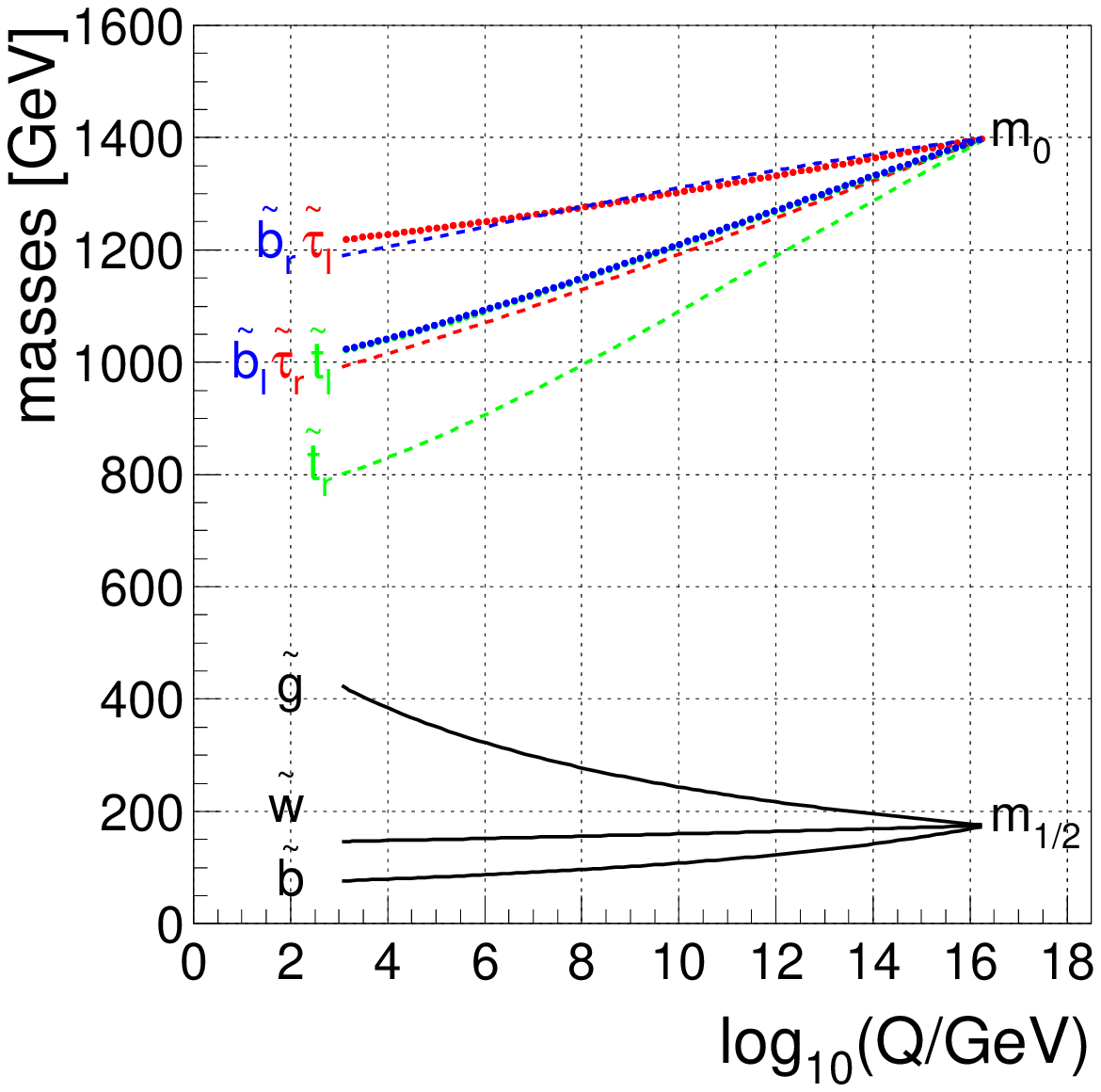}
 \caption[]{
 The light shaded (blue)  region in the $\mzero,\mhalf$ plane
  allowed by the combination of WMAP, EGRET, $b\rightarrow s\gamma$
 and the Higgs limit
for $\tb=51$, $\mu>0$ and $A_0=0$.
  The excluded regions, where the stau would be the LSP or EWSB fails or
 are indicated by the dots.
 }
 \label{relic}
\end{center}
\end{figure}
The lightest supersymmetric particle (LSP) is stable, if the
multiplicative quantum number R-parity, which is +1 for SM particles
and -1 for SUSY particles, is conserved. Non-conservation of
R-parity would lead to rapid proton decay\cite{susyrev}. The LSP is
a perfect candidate for Dark Matter and it can self annihilate into
fermion-antifermion pairs by Higgs or Z-exchange in the s-channel or
sfermion, chargino and neutralino exchange in the t-channel. The
dominant first three possibilities have amplitudes proportional to
the fermion mass, so heavy final states are preferred. For  values
of $\tb\approx 50$ the annihilation cross sections into
$b\overline{b}$ quarks are indeed of the order of magnitude required
by WMAP. For $m_{1/2}$ below 200 GeV, as required by the EGRET data,
the scalar masses have to be in the TeV range for a Higgs mass above
114 GeV, as shown in Fig. \ref{relic}. This region happens to
correspond to a thermally averaged annihilation cross section
$\sigma v\approx 2. 10^{-26} ~cm^3/s$, as required by the relic
density from WMAP. A typical mSUGRA spectrum is shown on the right
hand side of Fig. \ref{relic}. If $m_0$ is small compared with
$m_{1/2}$ the lightest lepton (usually the stau) can be lighter than
the neutralino, which happens in the left top corner on the left
hand side of Fig. \ref{relic}. In this case the LSP is charged and
cannot be DM. In the right bottom corner electroweak symmetry
breaking does not work.

\section{Objections to the DMA interpretation}\label{objections}

The DMA interpretation of the EGRET excess would mean that DM is not
so dark anymore, but DM is visible from the 30-40 flashes of
energetic gamma rays for each annihilation. This would be great, but
are there more mundane explanations? Attempts to modify the electron
and proton spectra from the locally measured spectra do not describe
the shape of the EGRET data in all sky directions,  as discussed in
detail before by comparing the EGRET data with the ``optimized
model''. Here we summarize some  other  (wrong) objections.

\begin{enumerate}
\item
Are the EGRET data reliable enough to make such strong conclusions?
The EGRET detector was calibrated in a quasi mono-energetic gamma
ray beam at the SLAC accelerator, so its response is well
known\cite{egret_cal}. Also the monitoring during the flight was
done carefully \cite{egret_cal1}. We have only used data in the
energy range between 0.1 and 10 GeV, where the efficiency is more or
less flat. So  the 9 years flight provided accurate and reliable
data, especially it would be hard to believe in an undetected
calibration problem, which would only effect the data above 0.5 GeV
and fake the gamma ray spectrum from the fragmentation of
mono-energetic quarks.
\item
The gamma ray spectrum above 0.1 GeV is dominated by pp-interactions
and therefore strongly dependent on the proton energy spectrum. This
cosmic ray spectrum
 was measured only locally in the solar neighbourhood. Could a harder spectrum
 near the Galactic center, where protons can be accelerated by the many supernovae
 there, explain the EGRET excess? No, first of all the energy loss times of
 protons with energies above a few GeV are longer than the lifetime of the universe,
 so by diffusion one expects
 everywhere the same energy spectrum. This is proven by the fact, that
 gamma ray spectrum for the Galactic center and the Galactic
 anticenter can be described by the {\it same} background shape.
\item
 Is the background well enough known to provide evidence for DMA?
The background is dominated by pp-collisions with a known shape and
fitting the normalization yields a ``self-calibrating'' background:
changing the slope of the proton spectra yields too many low energy
gamma rays, thus providing a shape different from the data. Fitting
this ``wrong'' shape with a free normalization reduces then the low
energy excess again and recovers the high energy excess. Note that
this ``self-calibration'' of the background also takes care of gas
clouds, ringlike or asymmetric structures in the background,
uncertainties in the absolute value of the total cross sections,
etc.
\item
  Is it possible to explain the excess in diffuse gamma rays with
unresolved point sources? This is unlikely, first of all since the
known point sources \cite{egret_cat} are only a small fraction of
the diffuse gamma rays and the majority of the resolved sources has
a rather soft spectrum, typically well below 1 GeV, as can be seen
from the plots in the Appendix. If this part of the spectrum would
be dominated by unresolved sources, then the diffuse component below
1 GeV would be lower than assumed, which in turn would lead to a
lower normalization of the background and a correspondingly stronger
excess for a fixed background shape. So arguing against DMA by
unresolved sources goes in the wrong direction.
\item
 Is one not over-interpreting the EGRET data by fitting so many
parameters for the different components: triaxial halo, inner ring
and outer ring? No, first of all the excess and enhancement in a
ringlike structure at 14 kpc was already discovered in the original
paper by \cite{hunter}. What we did is just trying to see if the
excess fits: a)  a single WIMP mass in all directions; b)  an
isothermal DM profile plus the substructure; c) the Galactic
rotation curve.
 Since each sky direction has 7 independent data points times 180 sky directions,
one has more than 1200 independent data points with the different DM
halo components being determined by independent sky directions: the
outer ring parameters are determined mainly by  30 sky directions
towards the Galactic anticenter, the inner ring parameters by ca. 15
sky directions towards the Galactic center and the triaxial halo
parameters by ca. 130 sky directions out of the Galactic plane. And
the most remarkable thing is that all these independent sky
directions all show an excess, which can be explained by a single
WIMP mass around 60 GeV. This is like having 180 independent
experiments at an accelerator all saying we see a significant excess
of gamma rays corresponding to $\pi^0$ production from
mono-energetic quarks. Then asked what mass they need to describe
the excess, they all say 60 GeV!
\item
 The tracing of DM relies largely on the outer rotation curve of
our Galaxy, which has  large uncertainties from the distance $r_0$
between the Sun and the Galactic center and is determined with a
different method than the inner rotation curve. Can this fake the
results? The outer rotation curve  indeed depends strongly on $r_0$,
as shown by \cite{honma}, who varied $r_0$ between 7 and 8.5 kpc. At
present one knows from the kinematics of the stars near the black
hole at the center of our Galaxy that $r_0=8\pm 0.4$ kpc
 \cite{rnull}, so the distance is already reasonably well known. But
whatever the value of $r_0$, the change in slope around 1.3$r_0$ is
always present, indicating a ringlike DM structure is always needed.
Furthermore the outer rotation curve shows first the same decrease
as the inner rotation curve and only then changes the slope, so the
different methods  agree between $r_0$ and 1.3$r_0$.
\item
 The outer ring at 14 kpc has a mass around  $9\cdot10^{10}$ solar
masses. This is around 50\% of the total mass inside the ring and
one may worry about the disk stability of the Milky Way by the
infall of such a heavy Galaxy. However, large spiral galaxies show
bumps of similar size \cite{sofue1}, so it seems not to be uncommon
to have masses of this size forming ringlike structures. Note that
only ringlike structures can form maxima and minima in the outer
rotation curve, since the rotation velocity squared  is proportional
to the {\it derivative} of the gravitational potential. Furthermore,
the stars in the ring at 14 kpc are all old, so the infall might
have happened before the growing of the disk.
\item
 One observes a ring of molecular hydrogen near the inner ring and
a ring of atomic hydrogen near the outer ring. Could this excess of
hydrogen not be responsible for the excess of the gamma rays? No,
our method  of fitting only the shapes with a free normalization
implies that this analysis is insensitive to density fluctuations of
the background, which change the normalization, not the shape.
\item
 How can one be sure that the outer ring originated from the tidal
disruption of a rather massive satellite galaxy, so one can expect
an enhanced DM density in the ring? One finds three independent
ringlike structures: stars, atomic hydrogen gas and excess of gamma
radiation. The stars show a scale height of several kpc and a low
velocity dispersion, so they cannot be part of the Galactic disk.
Therefore the infall of a satellite galaxy is the natural
explanation. Since the tidal forces are proportional to $1/r^3$, the
satellite will be disrupted most strongly at its pericenter, which
can lead to  DM density enhancements  at the pericenter after a few
orbits \cite{hayashi}. Some of the stars and gas may be caught in
this potential well. All three are found at 14 kpc with the stars
all being old  and more than 90\% of the mass being DM as deduced
from the strong EGRET excess at this radius.
\item
 Is it not peculiar that if a ringlike structure originates from
the infall of satellite galaxy, that it lies in the plane of the
Galaxy? No, in principle the infall can happen in all directions
with respect to the plane, but  the angular momenta of the inner
halo and a baryonic disk tend to align after a certain time by tidal
torques \cite{bailin}.
\end{enumerate}
\section{Summary and Outlook}
As mentioned in the Introduction, if DM is a thermal relic from the
early Universe, then it is known to annihilate, since the small
amount of relic density measured nowadays requires a large reduction
in its number density.  The annihilation into quark pairs will
produce $\pi^0$ mesons during the fragmentation into hadrons, which
in turn will decay into gamma rays.  For heavy WIMP masses the gamma
spectrum is considerably harder than the background spectrum. Such
an excess of hard gamma rays has indeed been observed by the EGRET
satellite and the relative contributions from background and DM
annihilation signal can be obtained by fitting their different
shapes with a free normalization factor for background and signal.
This method of a ``self-calibrating'' background yields results
practically independent of propagation models of our Galaxy.

This excess of hard diffuse gamma rays  shows all the features
expected from Dark Matter Annihilation:
\begin{itemize}
 \item The excess  is visible in {\it all} sky direction with a {\it
 same} spectrum   corresponding to the annihilation of
 WIMP pairs with a WIMP mass around 60 GeV.
\item The excess follows the distribution of a pseudo-isothermal
halo profile - as observed in many galaxies - with an additional
increased intensity at two toroidal shaped structures at radii of 14
and 4 kpc from the centre of the Galaxy. At these radii one finds an
enhanced density of hydrogen gas. Assuming that the baryons follow
the gravitational potential of DM one expects DM there. In addition,
at 14 kpc one has observed a ring of stars thought to originate from
the infall of a dwarf galaxy, while at 4 kpc one finds an enhanced
concentration of dust and molecular hydrogen. The dust shields the
latter against dissociation by UV radiation and can be collected in
the gravitational potential well  of a ring of DM.
\item Knowing the
halo profile of the DM together with the distribution of visible
matter allows to reconstruct the rotation curve of our Galaxy. The
gravitational potential wells of the ringlike DM substructure cause
minima and maxima in the rotation curve, since the rotation velocity
squared is proportional to the derivative of the gravitational
potential. These have indeed been observed and are perfectly
explained by the EGRET excess of gamma rays, thus proving that this
excess traces the DM in our Galaxy!
\end{itemize}

The results mentioned above make no assumption on the nature of the
Dark Matter, except that its annihilation produces hard gamma rays
consistent with the fragmentation of mono-energetic quarks between
50 and 100 GeV. WIMP masses in this range and the observed WIMP self
annihilation cross section  are consistent with WIMPs being the
Lightest Supersymmetric Particle predicted in the Minimal
Supersymmetric Model with supergravity inspired symmetry breaking,
called the mSUGRA model. The statistical significance of the EGRET
excess of at least 10 $\sigma$ combined with  all   features
mentioned above provides an intriguing hint that DM is not so dark,
but visible by its annihilation.

\acknowledgments
 I thank my close collaborators A. Gladyshev, D.
Kazakov, C. Sander and V. Zhukov for their contributions and V.
Moskalenko, A. Strong and  O. Reimer for numerous discussions on
galactic gamma rays and  the analysis of EGRET data.
 This work was supported by the BMBF (Bundesministerium f\"ur Bildung und Forschung) via the
 DLR.
%(Deutsches Zentrum f\"ur Luft- und Raumfahrt).


\begin{thebibliography}{99}
%
\bibitem{wmap} D.N. Spergel et al, 2003, ApJS, 148, 175;\\
C.L. Bennett et al., 2003, ApJS, 148, 1; See also:
http://map.gsfc.nasa.gov/m\_mm/pub\_papers/firstyear.html
\bibitem{jungman}
G.~Jungman, M.~Kamionkowski and K.~Griest,
%``Supersymmetric dark matter,''
Phys.\ Rep.\ {\bf 267} (1996) 195.\bibitem{bergstrom}
L.~Bergstr$\rm\ddot{o}$m,
%``Non-baryonic dark matter: Observational evidence and detection methods,''
Rept.\ Prog.\ Phys.\ {\bf 63} (2000) 793 [arXiv:hep-ph/0002126].
%%CITATION = HEP-PH 0002126;%%
\bibitem{sumner} Sumner, T.J.,
www.livingreviews.org/Articles/Volume5/2002-4sumner; Living Reviews
in Relativity published by the Max Planck Institute for
Gravitational Physics, Albert Einstein Institute, Germany.
\bibitem{bertone}
G.~Bertone, D.~Hooper and J.~Silk,
%``Particle dark matter: Evidence, candidates and constraints,''
arXiv:hep-ph/0404175.
\bibitem{deboer}
W.~de Boer, M.~Herold, C.~Sander, V.~Zhukov, A.~V.~Gladyshev and
D.~I.~Kazakov,
%``Excess of EGRET galactic gamma ray data interpreted
%as dark matter annihilation,''
arXiv:astro-ph/0408272.
\bibitem{deboer1}
W.~de Boer,
%``Evidence for dark matter annihilation from galactic
%gamma rays?,''
New Astron. Rev., 49, 213; arXiv:hep-ph/0408166.
%
\bibitem{pythia}
T. Sj\"ostrand, P. Eden, C. Friberg, L. L\"onnblad, G. Miu, S.
Mrenna and E. Norrbin, Computer Phys. Commun. {\bf 135} (2001) 238.
%(LU TP 00-30, hep-ph/0010017)
\bibitem{hunter}
 S.D. Hunter  { et al.},
 Astrophysical Journal {\bf 481}, 205 (1997).
\bibitem{optimized}
A.~W.~Strong, I.~V.~Moskalenko and O.~Reimer,
%``Diffuse Galactic
%continuum gamma rays. A model compatible with EGRET data and
%cosmic-ray measurements,''
Astrophys.\ J.\  {\bf 613}, 962 (2004); [arXiv:astro-ph/0406254].
\bibitem{kamae}
T.~Kamae, T.~Abe and T.~Koi,
%``Diffractive interaction and scaling
%violation in p p $\to$ pi0 interaction and GeV excess in galactic
%diffuse gamma-ray spectrum of EGRET,''
arXiv:astro-ph/0410617.
\bibitem{jimenez} R. Jimenez, L. Verde, and S.P. Oh,
         %``Dark halo properties from rotation curves,''
         MNRAS, {\bf 339} (2003) 243; arXiv:astro-ph/0201352.
\bibitem{nfw} N.F. Navarro, C.S. Frenk,  and S.D. White,
         %``A Universal Density Profile from Hierarchical Clustering,''
         ApJ {\bf 490}(1996) 493; arXiv:astro-ph/9611107.
\bibitem{newberg} H.J. Newberg,  et al., ApJ {\bf 569}(2002) 245.
\bibitem{ibata}
R.~A.~Ibata, M.~J.~Irwin, G.~F.~Lewis, A.~M.~N.~Ferguson and
N.~Tanvir,
%``One Ring to Encompass them All: A giant stellar structure that surrounds the
%Galaxy,''
Mon.\ Not.\ Roy.\ Astron.\ Soc.\  {\bf 340} (2003) L21;
[arXiv:astro-ph/0301067].
\bibitem{yanny}
B.~Yanny {\it et al.},
%``A Low Latitude Halo Stream around the Milky Way,''
Astrophys.\ J.\ {\bf 588} (2003) 824 [Erratum-ibid.\  {\bf 605}
(2004) 575]; [arXiv:astro-ph/0301029].
\bibitem{rocha-pinto1} J.D. Crane et al.
         %``Exploring Halo Substructure with Giant Stars: Spectroscopy of Stars in
         %the Galactic Anticenter Stellar Structure,''
         ApJ {\bf 594} (2003) L119; arXiv:astro-vph/0307505.
\bibitem{susyrev}
{\rm Reviews and original references can be found in:
 W.~de Boer,
%``Grand unified theories and supersymmetry in particle physics and cosmology,''%
 Prog.\ Part.\ Nucl.\ Phys.\ {\bf 33} (1994) 201 [arXiv:hep-ph/9402266];
 \\
 H.E. Haber, Lectures
 given at Theoretical Advanced Study Institute, University of Colorado, June
 1992, Preprint Univ. of Sante Cruz, SCIPP 92/33; see also SCIPP 93/22;\\ {\it
 Perspectives on Higgs Physics}, G. Kane (Ed.), World Scientific, Singapore
 (1993);\\
% {\it Int. Workshop on Supersymmetry and Unification}, P. Nath
 %(Ed.), World Scientific, Singapore (1993);\\ {\it Phenomenological Aspects of
 %Supersymmetry}, W. Hollik, R. R\"uckl and J. Wess (Eds.), Springer Verlag
 %(1993);\\ R. Barbieri, Riv. Nuovo Cim. {\bf 11} (1988) 1;\\
 A.B. Lahanus and
 D.V. Nanopoulos, Phys. Rep. {\bf 145} (1987) 1;\\ H.E. Haber and G.L. Kane,
 Phys. Rep. {\bf 117} (1985) 75;\\ M.F. Sohnius, Phys. Rep. {\bf 128} (1985)
 39;\\ H.P. Nilles, Phys. Rep. {\bf 110} (1984) 1;\\ P. Fayet and S. Ferrara,
 Phys. Rep. {\bf 32} (1977) 249.}
%
\bibitem{bs}
W.~de Boer and C.~Sander,
%``Global Electroweak Fits and Gauge Coupling Unification,''
arXiv:hep-ph/0307049 and references therein.
\bibitem{egret_cal} Thompson D.J. et al.,
         %``Calibration of the EGRET high-energy gamma-ray telescope in the range
         %20-MeV to 10000-MeV with a tunable beam of quasi-monoenergetic gamma-rays at SLAC,''
         IEEE Trans. Nucl. Sci., {\bf 34}(1987) 36.
\bibitem{egret_cal1} J.A. Esposito et al.,
         %``In-Flight Calibration of EGRET on the Compton Gamma-Ray Observatory''
         ApJ {\bf 123}(1987)  203.
\bibitem{egret_cat}
        R.C. Hartman,  {\it et al.},
        %``The Third EGRET catalog of high-energy gamma-ray sources,''
        ApJ {\bf S123}(1999)  79.
\bibitem{honma}
        M.~Honma and Y.~Sofue,
        %``Mass of the Galaxy inferred from outer rotation curve,''
        Publ. of the Astronomical Society of Japan, v.48, p.L103-L106;
        arXiv:astro-ph/9611156.
\bibitem{rnull}
        F.  Eisenhauer,  {\it et al.},
        %``A Geometric Determination of the Distance to the Galactic Center,''
        ApJ. {\bf  597} (2003)  L121; arXiv:astro-ph/0306220.
\bibitem{sofue1} Y. Sofue,
        %``Central Rotation Curves of Galaxies''
%        in "Galaxy Disks and Disk Galaxies", ASP Conf. Series, Vatical Conf. Rome 2000,
%        eds. J. Funes, and E.M. Corsini,
        arXiv:astro-ph/0010595.
\bibitem{hayashi}
        E. Hayashi, et al.,
        %``The Structural Evolution of Substructure,''
        ApJ  {\bf 584}(2002) 541; arXiv:astro-ph/0203004.
\bibitem{bailin}
          J. Bailin, J. { et al.},
          %``Internal Alignment of the Halos of Disk Galaxies in Cosmological
          %Hydrodynamic Simulations,''
          arXiv:astro-ph/0505523.
\end{thebibliography}
\end{document}